\documentclass[a4paper]{article}
\usepackage{authblk}
\usepackage[T1]{fontenc}
\usepackage{amsmath}
\usepackage{amssymb}
\usepackage{bm}
\usepackage{graphicx}
\usepackage{hyperref}
\usepackage{bbm}

\usepackage{algorithm}
\usepackage{algpseudocode}

\usepackage{float}

\newcommand{\UV}[1]{\hat{\boldsymbol{#1}}}

\begin{document}

\title{Approximate method for helical particle trajectory reconstruction in high energy physics experiments}

\author[1]{K. Topolnicki \thanks{kacper.topolnicki@uj.edu.pl}} 
\author[2]{T. Bold \thanks{tomasz.bold@cern.ch}}
\affil[1]{M. Smoluchowski Institute of Physics, Jagiellonian University 30-348, Kraków, Poland}
\affil[2]{AGH University of Science And Technology, Dept. of Physics and Applied Computer Science, Krakow, Poland}

\maketitle

\begin{abstract}

High energy physics experiments, in particular experiments at the LHC, require the reconstruction of charged particle trajectories. Methods of reconstructing such trajectories have been known for decades, yet the applications at High Luminosity LHC require this reconstruction to be fast enough to be suitable for online event filtering. 

A particle traversing the detector volume leaves signals in active detector elements from which the trajectory is reconstructed. If the detector is submerged in a uniform magnetic field that trajectory is approximately helical. Since a collision event results in the production of many particles, especially at high luminosities, the first phase of trajectory reconstruction is the formation of candidate trajectories composed of a small subset of detector measurements that are then subject of resource intensive precise track parameters estimation.

In this paper, we suggest a new approach that could be used to perform this classification. The proposed procedure
utilizes the $z$ coordinate in the longitudinal direction
in addition to the $x, y$ coordinates in the plane perpendicular to the direction of the magnetic field. The suggested algorithm  works equally well
for helical trajectories with different proximities to the
beamline which is beneficial when searching for  products of particles
with longer lifetimes.

\end{abstract}

\section{Introduction}

Due to the enormous QCD cross-section at hadron colliders, processes of interest often occur with significant background. For instance, at the High Luminosity LHC~\cite{EPJ_WOC_2018}, about 200 parasitic collisions are expected to occur in every bunch crossing. Those additional collisions contribute signals to all detector elements. The most notable bias is incurred to the energy measured by the calorimeters.  The only experimental tool able to disentangle signals from the primary collision of interest from other collisions is precise charged particle tracking. The use of tracks is not limited to the reconstruction of recorded events for offline analysis but is also necessary when selecting events in online filters. 
Another application is triggering on tracking-only signatures.
Of utmost interest are signatures of long-lived particles which decay far from the collision region but typically have unspecific calorimetric signatures. Performing tracking capable of finding such types of tracks can significantly increase sensitivities and even enable searches for Beyond Standard Model Particles~\cite{Bobrovskyi2013}.

The track reconstruction is customarily divided into several stages: 
\begin{enumerate}
    \item detector data decoding, 
    \item searching for track candidates,  
    \item track fitting.
\end{enumerate}
The computational complexity of these stages is of a different character. The execution time of the first step is a linear function of the number of detector signals in an event. The goal of the second step is to select subset of signals that could constitute a charged particle track. Thus, this step is combinatorial by nature and thus has an undesired computational complexity dependence on the number of detected signals. In addition, the candidates found in this step are processed by the third, fitting step, and thus many false positives found at step 2 result in a large number of fits being performed at step 3.

In summary, for a time-constrained tracking system, an algorithm responsible for forming the track candidates that is fast, has high efficiency and does not produce an excessive number of spurious candidates is of key importance.


In preparation to the HL-LHC data acquisition conditions, experiments prepare tracking algorithms and systems for online filters~\cite{Collaboration_2008,Collaboration2_2017,Collaboration1_2017,ATLASTDAQAmendment}. A lot of effort is invested in studying compute accelerators like FPGAs or GPUs to more efficiently perform the algorithms that are already very well known~\cite{Bartz:2019dkp}. 
Novel algorithmic approaches are also investigated. For instance the ATLAS experiment attempts to use the Hough Transform (HT) algorithm~\cite{HT_1,HT_2,Novel,Rinaldi}. In the HT there is a trade-off between the computational complexity, which is linear as function number of the number of inputs, and the high memory consumption of the algorithm. With a simplifying assumption that the tracks originate from a known origin (primary particles) the resources in currently available hardware are sufficient for a well-performing implementation. Constraining the HT  to tracks of high momentum particles further simplifies the algorithm.
For a good overview of current classification methods, we refer the reader to~\cite{newbook} and references therein.

In this paper, we describe a fast algorithm to perform track candidates formations that is similar to HT yet is not limited to particles originating in the collision zone. Similar to HT, narrowing the application scope to high momentum particles results in numerical simplifications. 

\section{Algorithm overview} 

\begin{figure}[H]
	\centering
	\includegraphics[height = 0.8 \textheight]{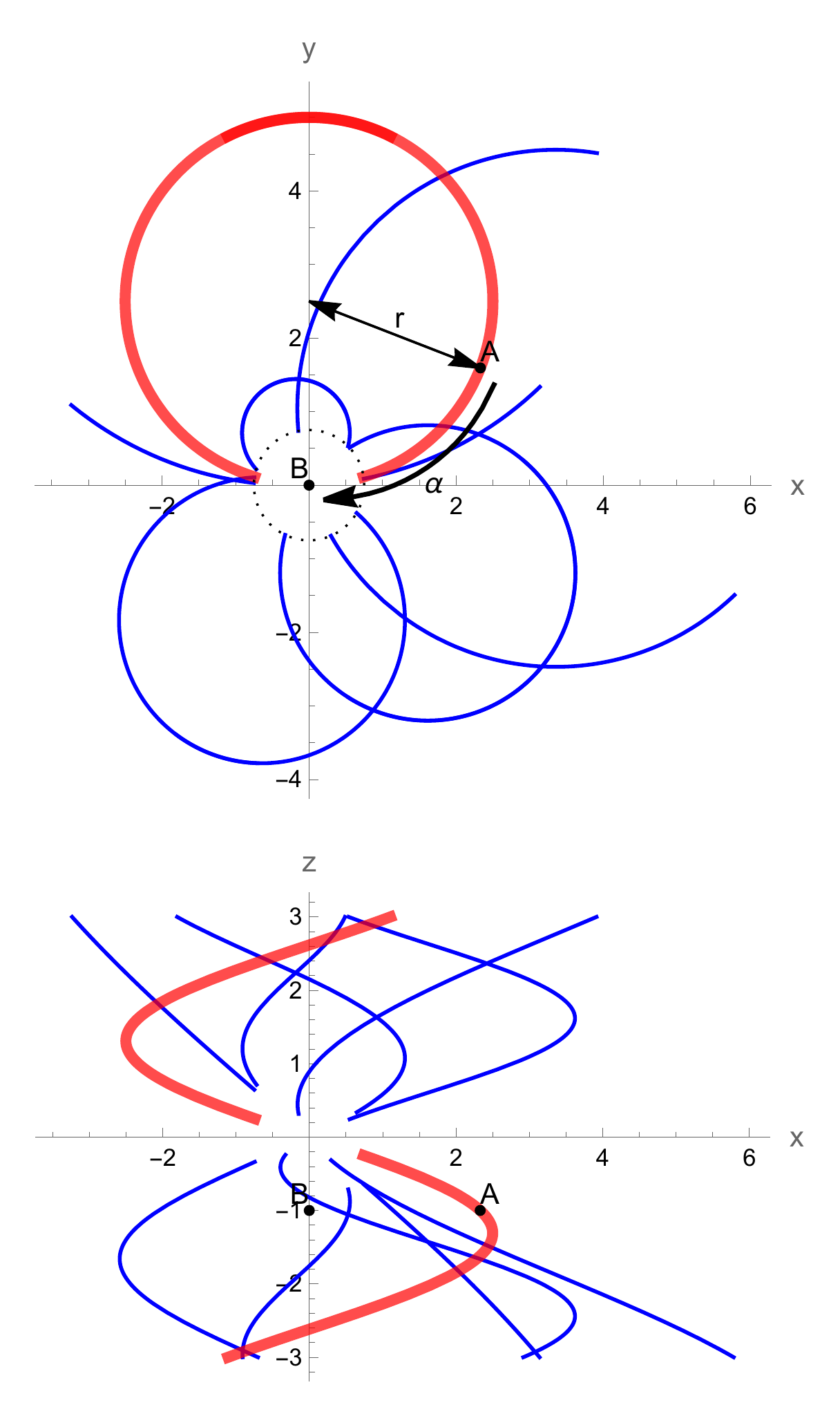}
	\caption{(color online) Helical particle tracks inside a detector. The magnetic field induction vector is aligned with the $\UV{z}$ direction.
	The $\UV{x}$ - $\UV{y}$ projection is shown on the top and the $\UV{x}$ - $\UV{z}$
	 on the bottom drawing. 
	One helix, with radius $r$, is selected and plotted in red. Rotating measurement $A$ on this helix by angle $\alpha$ round the $\UV{z}$ axis 
	with the center of rotation at the red helix center  
	will transform it into position $B$. A suitable choice of the $\alpha = \alpha(z)$ 
	dependence can transform all points lying on the red helix into a straight line in the direction of $\UV{z}$. 
	Data in a cylinder around the $\UV{z}$ axis was
	removed to reduce background noise, this is illustrated using the dotted
	circle on the top plot.}
	\label{situation_before}
\end{figure}

A diagrammatic illustration of the charged particle trajectories in a uniform magnetic field is shown in Fig. \ref{situation_before}.  
Before colliding, particles are assumed to travel along the $\UV{z}$ axis and therefore particle tracks start close to the origin of the $\UV{x} - \UV{y}$ coordinate system.
Sections of particle tracks within a small cylinder around $\UV{z}$ were removed. This is illustrated with a
dotted circle in the $\UV{x} - \UV{y}$ projection and it reflects the experimental reality in which detectors are absent in the vicinity of beams. 
Coincidentally, it has practical implications in the algorithm that result in the reduction of spurious candidates.
If the magnetic field is assumed to be in the $\UV{z}$ direction, along the beam-line, 
then the charged particles will travel in helical trajectories with helix axis along $\UV{z}$; these are marked in blue and red. 
It should be noted that the starting 
position of the particle tracks in any of the $\UV{z}$, $\UV{x}$ and $\UV{y}$ directions is not relevant to the suggested algorithm.

\begin{figure}[H]
	\centering
	\includegraphics[width = 0.5 \textwidth]{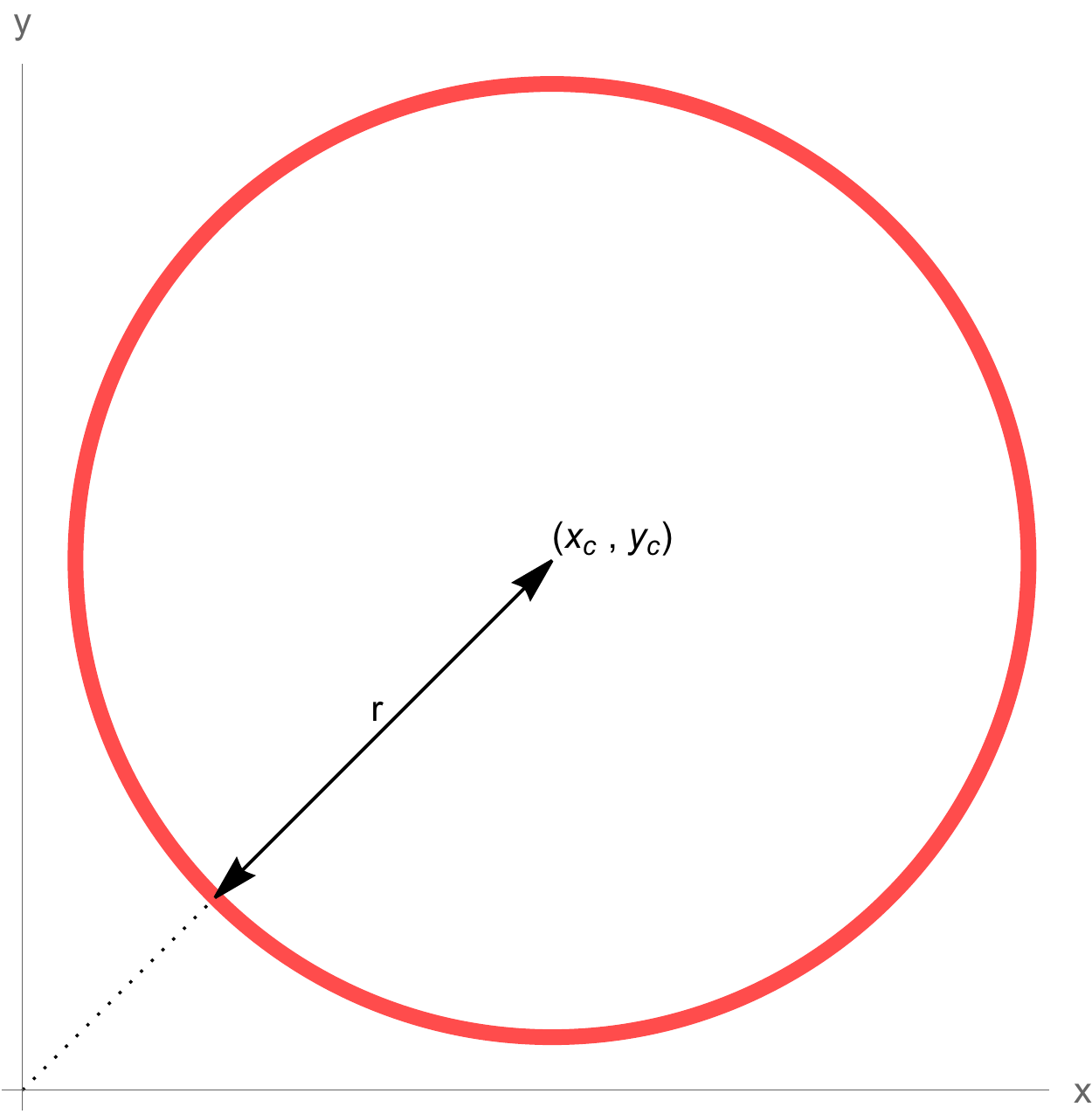}
	\caption{(color online) The visualisation of helix  
	parameterisation by the $z$ coordinate and given explicitly by:
	$(x_{c} , y_{c} , 0) + 
	\left( r cos(-\frac{\nu (z - z_{0})}{r}) , r sin(-\frac{\nu (z - z_{0})}{r}) , z \right)$.
	The helix radius is $r$
	and the helix axis is at position $(x_{c} , y_{c})$ in the $\UV{x}-\UV{y}$ plane. 
	The $\nu$ parameter determines the pitch of the helix. 
	Changing $z_{0}$ moves the helix up or down along the $\UV{z}$ axis (rotating the helix about its center has the same effect). 
	}
	\label{parametrization}
\end{figure}

As is customary for track seeding, the discussed algorithm relies on the unbound measurements, that is absolute positions of measurement points. The detector registers a set of $N$ measurements:
\begin{equation} 
	D = \{ (x_{i} , y_{i} , z_{i}) , i = 1 \ldots N\}.
	\label{d}
\end{equation}
The goal is to classify which measurements can be attributed
to the same helix. To solve this problem, we will be discussing a carefully chosen function 
$u_{x_{c} , y_{c} , \nu}(x , y , z)$
where $x_{c} , y_{c} , \nu$ are helix parameters illustrated and described in 
Fig. \ref{parametrization}. 
Mapping this function over $D$ will produce a new set of three-dimensional points,
\begin{equation} 
	D' = \{ (x'_{i} , y'_{i} , z'_{i}) = u_{x_{c} , y_{c} , \nu}(x_{i} , y_{i} , z_{i}) , i = 1 \ldots N\}.
	\label{d_prime}
\end{equation}
that after binning in the $\UV{x} - \UV{y}$ plane can be used
to classify subsets of measurements to a helix with given parameters 
$x_{c} , y_{c} , \nu$. This two dimensional histogram can also be used to read off
the two remaining helix parameters $r$ and $z_{0}$ from Fig. \ref{parametrization}.

The suggested classification algorithm can be performed in a loop over the
three helix parameters $x_{c}$, $y_{c}$, and $\nu$. 
The execution of the main loop of the procedure 
can be efficiently parallelized since each
iteration is independent. 
In each iteration:
\begin{enumerate}
    \item The detected measurements $D$ from (\ref{d}) are transformed into $D'$ from (\ref{d_prime}) using $u_{x_{c} , y_{c} , \nu}(x , y , z)$.
    \item The measurements in $D'$ are binned in the $\UV{x} - \UV{y}$ plane to create a two dimensional histogram.
    \item If the helix parameters $x_{c} , y_{c} , \nu$ happen to match those of a real particle track inside the detector then an isolated and well distinguished peaks should be present in the histogram. 
    \begin{enumerate}
        \item The two remaining helix parameters $r$ and $z_{0}$ from Fig. \ref{parametrization} can be
            determined from the position of the peak on the histogram.
    \end{enumerate}
\end{enumerate}

The $u_{x_{c} , y_{c} , \nu}$ transformation will be discussed in the next section. The same form of $u_{x_{c} , y_{c} , \nu}$ 
works just as well for particle tracks that closely approach the beam line
and particle tracks that start at large distances from the beam.
This opens up
the possibility to detect particles that originate at large distances from the interaction region. 
Additionally, we consider a reversible transformation, no information is discarded and 
all three coordinates of each point in the set $D$ are utilized. 
This is different from some other methods, like the HT~\cite{HT_1,HT_2,Novel,Rinaldi}
that usually discard the $\UV{z}$ coordinate.

The paper is organized as follows. In the next section 
we introduce the 
transformation $u_{x_{c} , y_{c} , \nu}$ that will be used in the classification
algorithm. In section \ref{statistical_performance}, we employ a realistic
Monte Carlo simulation to determine the properties (efficiency, sensitivity, specificity, 
predictive value) of this new approach. Finally, section \ref{sec_summary_and_outlook}
contains the summary and outlook.

\section{Unraveling transformation}
\label{sec_unravel_transform}

The $u_{x_{c} , y_{c} , \nu}(x , y , z)$ transformation that we will be utilizing is illustrated in Fig. \ref{situation_before}.
For assumed helix parameters (red helix) it takes measurement $A$ and rotates it by an angle $\alpha$ around the helix axis. 
The resulting point $B$ lies outside of the helix as its $z$ coordinate remains unchanged. If the angle of the rotation, $\alpha$, 
depends on the $z$ coordinate
of a helix point according to $\alpha = \frac{z \nu}{r}$  then all measurements of the same (red) helical trajectory will be arranged along a straight line along the $\UV{z}$ axis in the set $D'$ from~\eqref{d_prime}. Such a transformation effectively ``unravels`` the helix into a straight line. 

\begin{figure}[H]
	\centering
	\includegraphics[height = 0.7 \textheight]{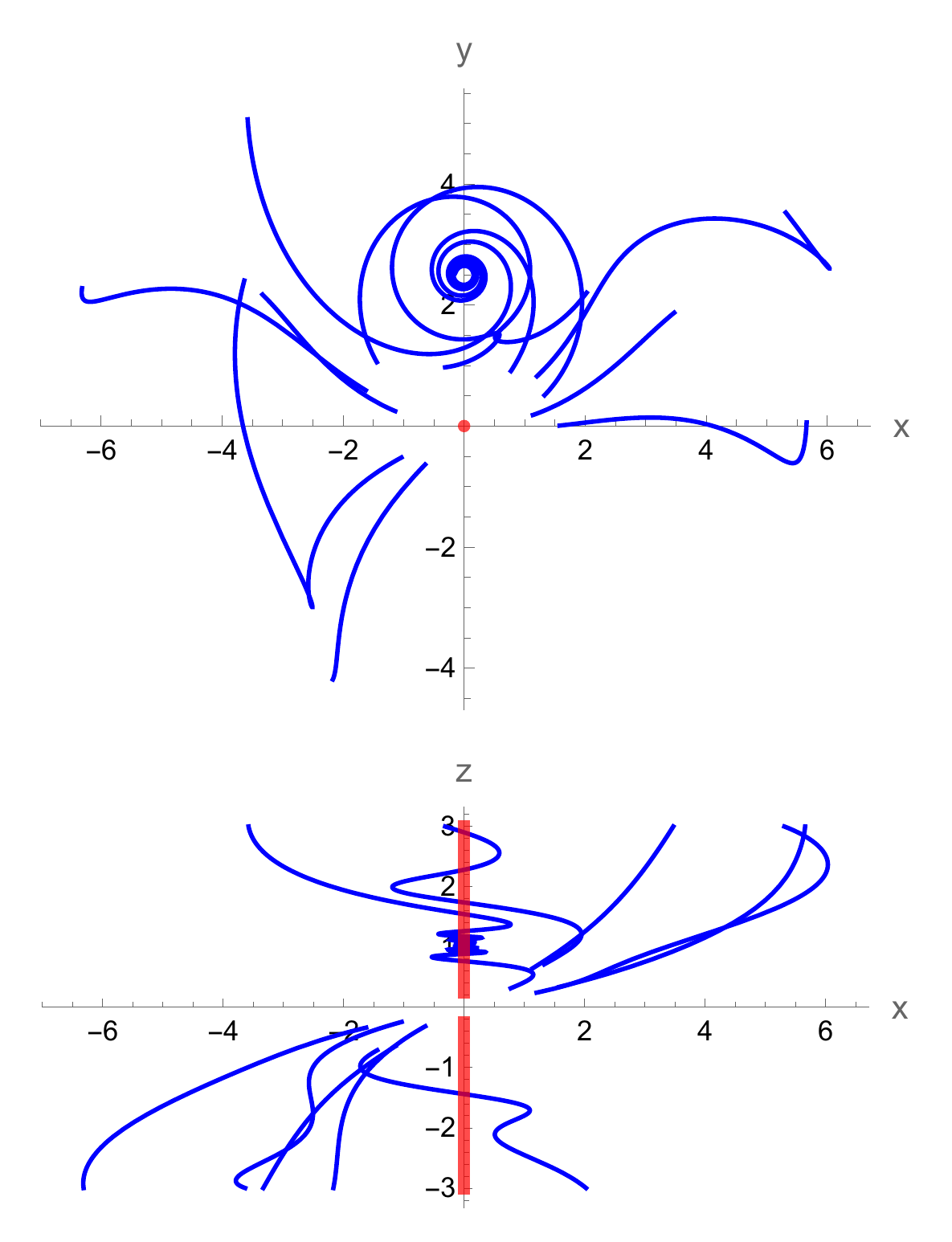}
	\caption{(color online) Similar to Fig. \ref{situation_before} but with transformation (\ref{u_transformation}) applied to each
	particle track. The $x_{0}$, $y_{0}$ and $\nu$ parameters were chosen such that the red helical path
	from Fig. \ref{situation_before} is transformed into a straight line along the $\UV{z}$ 
	direction.}
	\label{situation_after}
\end{figure}

The explicit form of the ``unraveling`` transformation is given by the equation:
\begin{align} \nonumber
	u_{x_{c} , y_{c} , \nu}(x , y , z) &:= 
	(x_{c} , y_{c} , 0) \\
	&+ R_{\UV{z}}\left(\frac{z \nu}{\sqrt{(x - x_{c})^{2} + (y - y_{c})^{2}}}\right)  
	\left( (x , y , z) - (x_{c} , y_{c} , 0) \right)
	\label{u_transformation}
\end{align}
where $R_{\UV{z}}(\alpha)$ is the rotation matrix by angle $\alpha$ around the $\UV{z}$ axis.
The result of applying (\ref{u_transformation}) to all helices from Fig. \ref{situation_before}
but with parameters $x_{c} , y_{c} , \nu$ tailored to the red helix can be seen
in Fig. \ref{situation_after}. 
The trajectory of interest (red) is ``unraveled'' and becomes the straight red line
(this is seen on the $\UV{x}$ - $\UV{z}$ projection and as a red dot on the $\UV{x} - \UV{y}$ projection) 
whereas helices with different parameters (blue) remain curved
after the transformation. In the $\UV{x}$ - $\UV{y}$ projection the helix of interest degenerates to a Dirac delta function in a single point at $(0,0)$ while the projections of other helices occupy the whole $\UV{x}$ - $\UV{y}$ space. With a finite number of measurements per helix, this will result in a peak being formed.
Additionally, if the algorithm application domain would be limited to high momentum particles with large helix radii then \eqref{u_transformation} would simplify to linear translations whose parameters depend on the distance 
from $(x_{c}, y_{c})$.

It is interesting to see what happens to a point on a helical particle track whose
Cartesian coordinates are $(x , y , 0)$ after the application of
\eqref{u_transformation}. Setting the value of $z = 0$
results in the rotation operator becoming and identity $R_{\UV{z}}(0) = \mathbbm{1}$ and:
\begin{equation*}
	u_{x_{c} , y_{c} , \nu}(x , y , 0) = (x , y , 0).
\end{equation*}
This means that if a peak (corresponding to a
helix with parameters $x_{c} , y_{c} , \nu$) is observed on the 
$\UV{x} - \UV{y}$ histogram of $D'$ from \eqref{d_prime} then the position of this peak 
on the $\UV{x} - \UV{y}$ plane will correspond to the particle 
trajectory at $z = 0$. The peak position can therefore be used to determine the
two remaining parameters $r$ and $z_{0}$ from Fig. \ref{parametrization}.

\begin{figure}[H]
	\centering
	\includegraphics[width = 1.0 \textwidth]{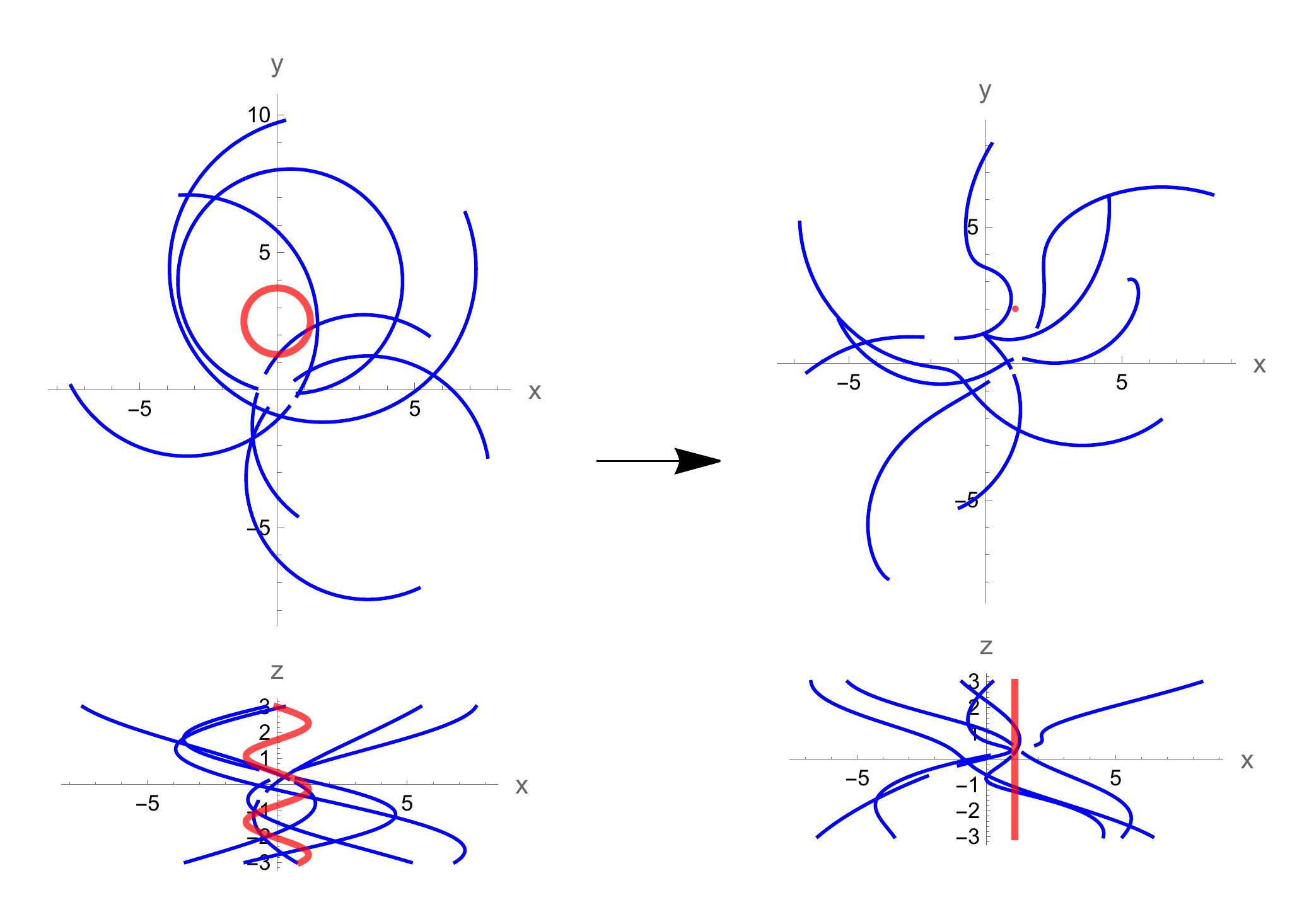} 
	\caption{(color online) The effect of applying the ``unraveling'' transformation to particle
	trajectories shifted away from the origin. The left diagram is similar to Fig. \ref{situation_before} and shows the situation before ``unraveling''. The right diagram 
	is similar to Fig. \ref{situation_after} and shows the situation after the ``unraveling'' transformation, with the red helix parameters, was applied.
	The red trajectory is projected into a point on the $\UV{x} - \UV{y}$
	distinguishing it from the other, blue, trajectories.
	}
	\label{prox}
\end{figure}

It is also important to note that the proximity of particle tracks to the beam-line does not change the suggested procedure. 
An example is illustrated in Fig.~\ref{prox}. Initially, the tracks on the left side of the figure,
are shifted away from the beam-line. The ``unraveling''
transformation, whose parameters are tailored to the specific (red) helix, turns
it into a straight line along $\UV{z}$ and on the $\UV{x} - \UV{y}$
plane  into a point that distinguishes
it from the other (blue) trajectories.

\section{Algorithm performance}
\label{statistical_performance}

In this section 
we present the performance of the aforementioned algorithm in terms of the efficiency / sensitivity and predictive value.
For that purpose a simulation a with relatively realistic Open Data Detector~\cite{ODD} is used. This detector features a standard design with pixel layers closer to beam and silicon strip detectors further away from the beam line. In the barrel region it consists of 4 layers of pixel detector in the innermost part, followed by 6 layers of strip modules. The endcap consists of 7 pixel and 6 strip layers perpendicular to the beam line. Fist the $p-p$ collisions at nominal LHC energies were generated using Pythia 8.2~\cite{pythia} and 200 of them were superimposed to simulate high pileup conditions. Then the detector response was simulated with the  FATRAS~\cite{FATRAS} that is part of ACTS framework~\cite{ACTS}. The particles have falling momentum spectra, cover the full azimuth and span 6 units of  pseudorapidity\footnote{Pseudorapidity, a measure of polar angle $\theta$ in cylindrical coordinate systems typically used in HEP is defined as: $\eta = -ln(tan(\theta/2))$.}.

To study the algorithm performance the sizes and shapes of the $\UV{x} - \UV{y}$ bins were established with relation to the  the resolution of the detector in~\ref{bin_shape}. 
In~\ref{binning_procedure} we describe an effective binning procedure
necessary to carry out  calculations.
Next, the statistical approach is described in~\ref{mc} and 
in~\ref{results} the results are shown. Finally section~\ref{step_size} contains some additional
considerations about optimal algorithm parameters for scanning a wide range of the helix parameter space.

\subsection{Bin shape and size}
\label{bin_shape}

\begin{figure}[H]
	\centering
	\includegraphics[width = 0.75 \textwidth]{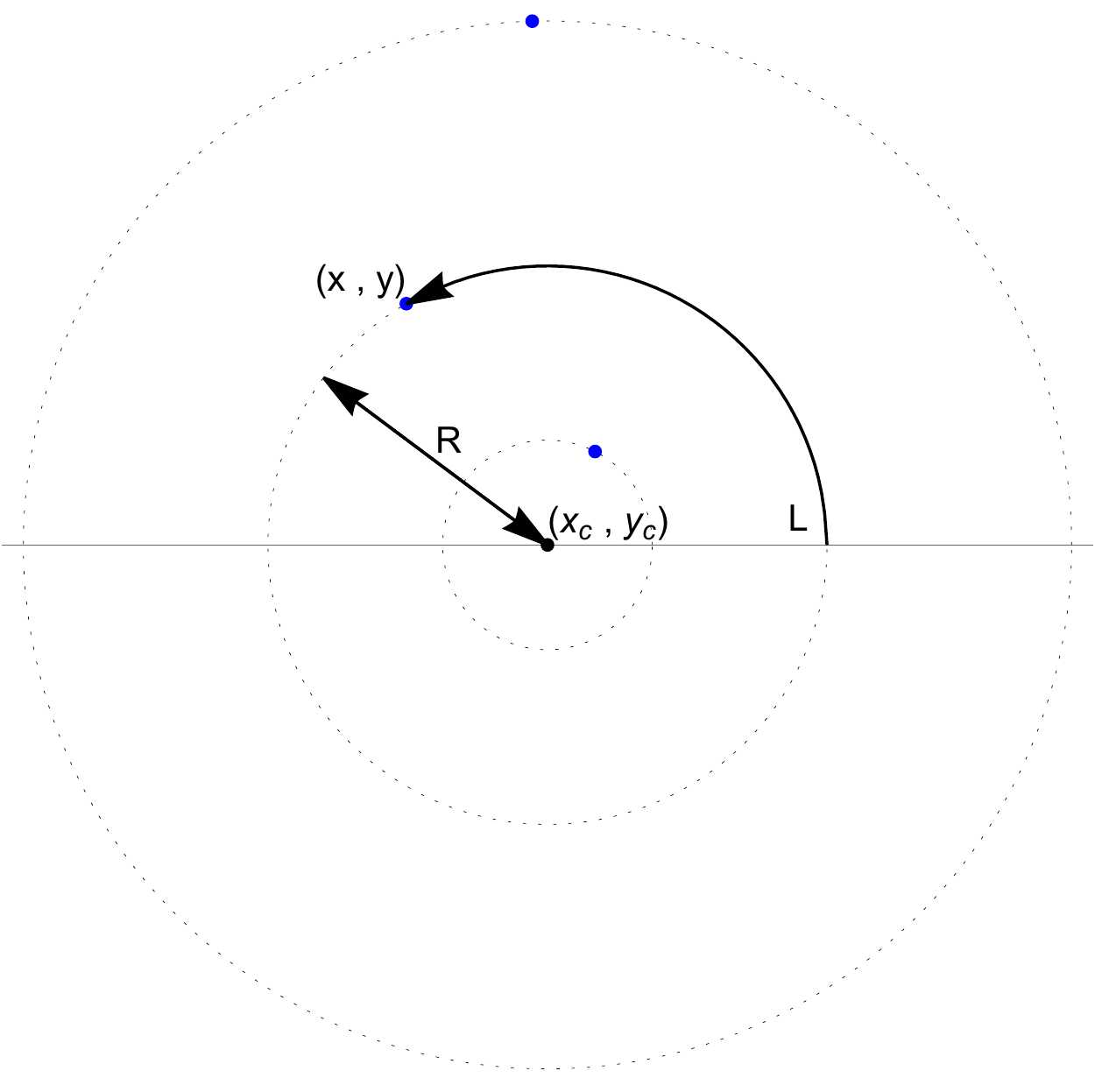} 
	\caption{(color online) Blue dots mark points from $D'$ projected onto the $\UV{x}-\UV{y}$ plane. 
	The distance $R$ of a point from the helix center $(x_{c} , y_{c})$ and the travel distance around 
	a potential helix projection $L$ are used for binning. 
	}
	\label{bin_coord}
\end{figure}

\begin{figure}[H]
	\centering
	\includegraphics[width = 0.75 \textwidth]{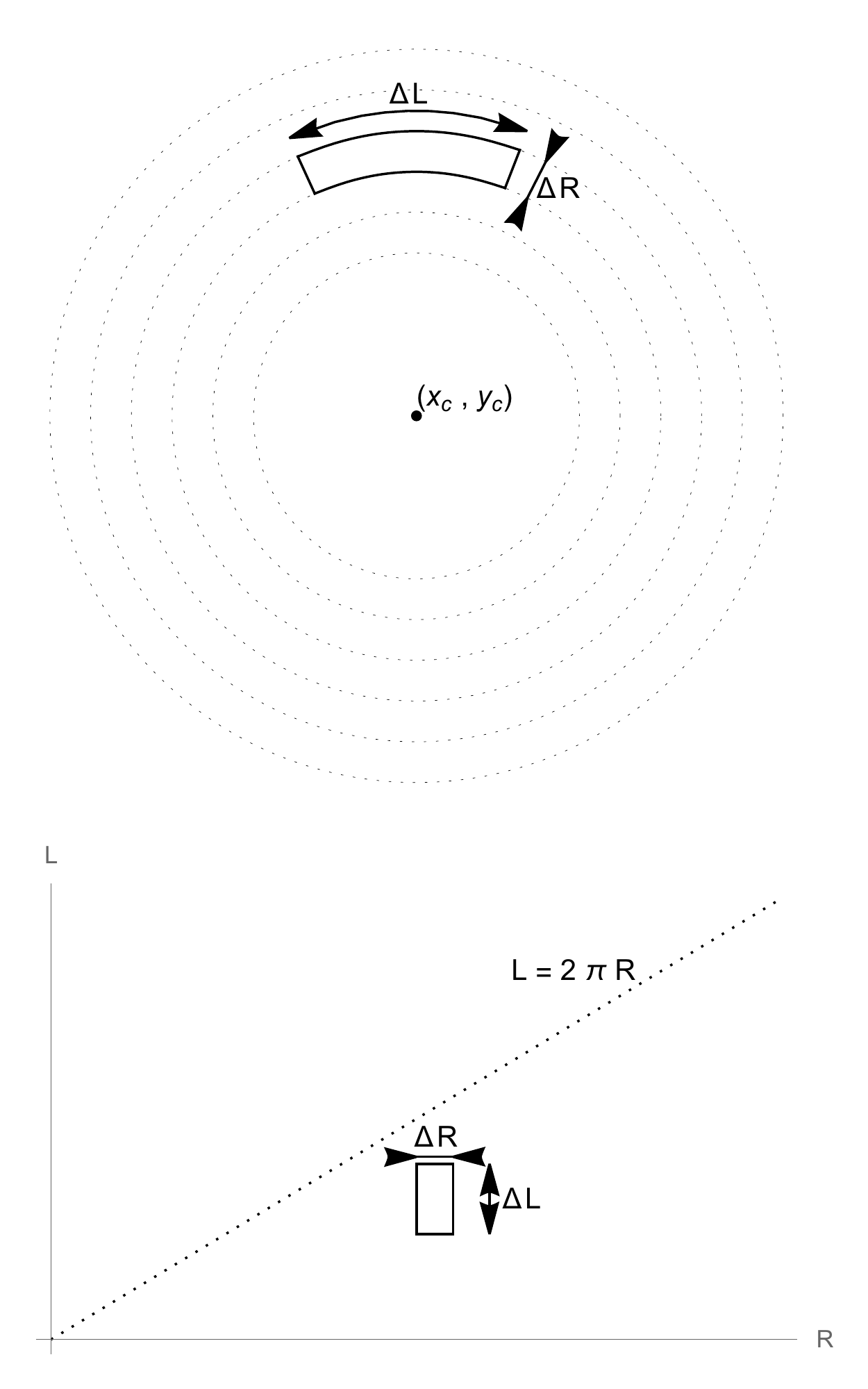} 
	\caption{The new coordinate system from Fig.~\ref{bin_coord} is used to form 
	rectangular $\Delta R \times \Delta L$ bins in the $\UV{R}-\UV{L}$ plane. Hits are expected on the lower part of this plane, the maximum value of $L$ for a given value of $R$
	is shown using the dotted line.  
	}
	\label{bins}
\end{figure}

In order to create the $\UV{x}-\UV{y}$ histograms in $D'$
we use a new coordinate system that
is illustrated in Fig.~\ref{bin_coord}. The distance $R$ of a point from the helix center $(x_{c} , y_{c})$ and the travel distance around a potential helix projection $L$ are used. The resulting rectangular bins
in the new $\UV{R}-\UV{L}$ plane are illustrated in Fig.~\ref{bins}. The size of the bins in the $\UV{R}$ and $\UV{L}$ directions are $\Delta R$ and $\Delta L$ respectively. 

\begin{figure}[H]
	\centering
	\includegraphics[width = 0.75 \textwidth]{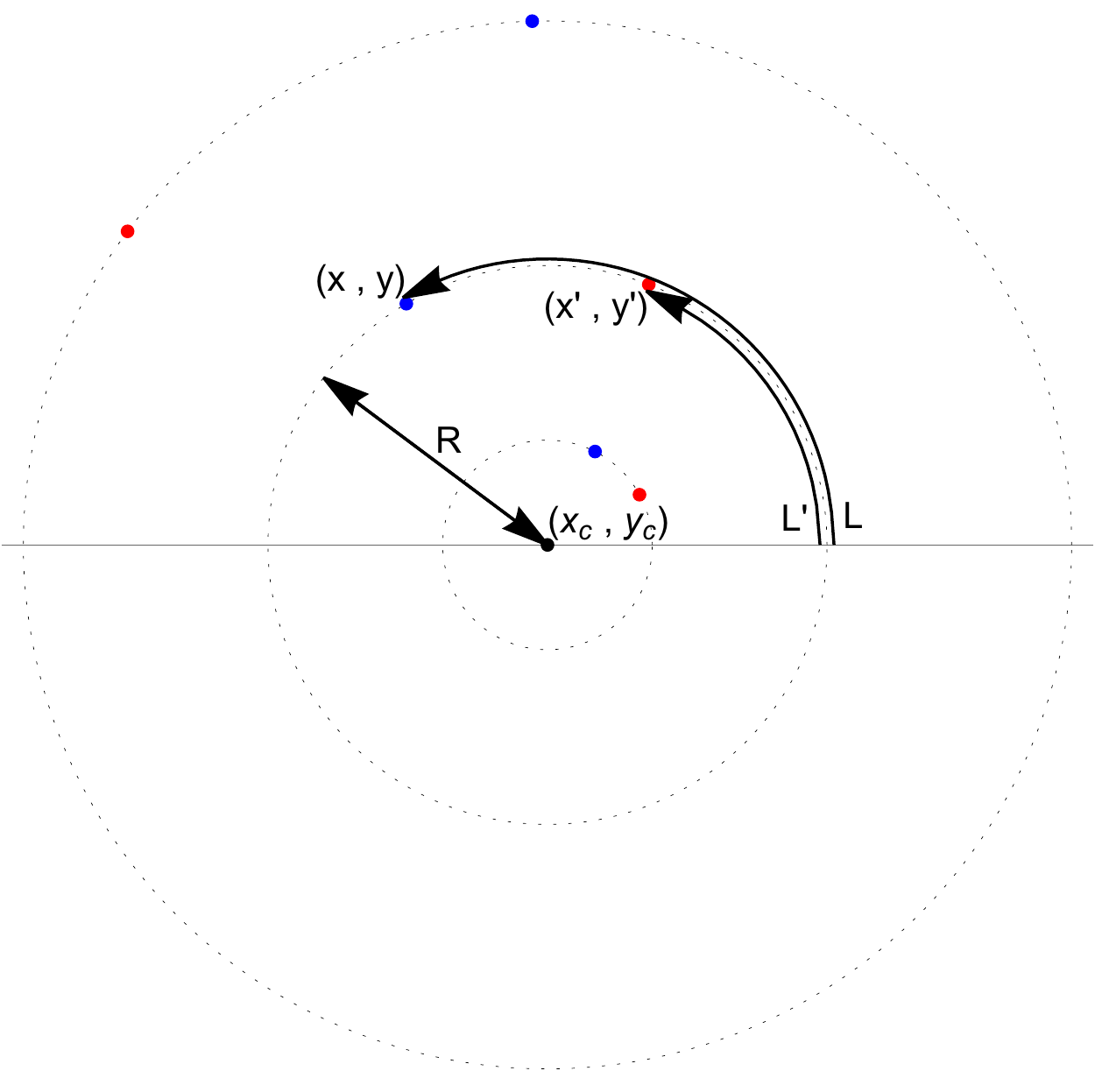} 
	\caption{(color online) The red and blue dots are points in $D'$ from (\ref{d_prime}). The distance $L' - L$
	results from the uncertainty $\sigma_{z}$ of $z$ coordinate measurements in $D$ from (\ref{d}) and is
	equal to $\nu \sigma_{z}$ where $\nu$ is the helix pitch (see Fig. \ref{parametrization}).
	}
	\label{bin_transform}
\end{figure}

The values of $\Delta R$ and $\Delta L$ should reflect the spatial resolution of a given detector.
Points from $D'$, illustrated using red and blue dots on Fig.~\ref{bin_transform}, can result from the same
measurements in $D$ if we assume that a spatial uncertainty of the measurement in the $\UV{z}$ direction,
$\sigma_{z}$, is present. The difference between $L$ and $L'$ from Fig.~\ref{bin_transform} will be proportional to this uncertainty multiplied by the helix pitch. Using this observation a good choice for the bin width in the $L$ direction is:
\begin{equation}
\Delta L = \nu \Delta_{z},
\label{bin_L}
\end{equation}
where $\Delta_{z}$ is a constant parameter, proportional to 
the uncertainty $\sigma_{z}$.
The bin size in $R$ is easier to estimate because (\ref{u_transformation}) does not change this value. 
$\Delta R$ will be directly proportional to the uncertainty of the measurement in the $\UV{x}-\UV{y}$ plane, that is in experimental reality a function of detector alignment and sensors resolution. This suggests a good choice for the bin width in the $R$ direction:
\begin{equation}
\Delta R = \Delta_{xy},
\label{bin_R}
\end{equation}
where $\Delta_{xy}$ is a constant parameter, proportional to $\sigma_{xy}$.
If $\Delta_{xy} \approx \sigma_{xy}$ and $\Delta_{z} \approx \sigma_{z}$ then we can expect most of the 
points from $D'$ that belong to a single chosen particle track to fall into a single rectangular
$\Delta L \times \Delta R$ bin if the parameters of the ``unraveling" transformation match those of the chosen helical trajectory.
The choice of $\Delta_{xy}$ and $\Delta_{z}$ has a big impact on the performance of the algorithm as shown in section~\ref{results}.

\subsection{Binning procedure}
\label{binning_procedure}

In order to conserve memory, we developed an efficient method to perform the binning. 
Memory consumption is precisely proportional to the number of measurements.
In the first
step the values of $R$ and $L$ are calculated for every point in $D'$. 
Next, knowing $\Delta_{R}$ and $\Delta_{L}$, the bin numbers $n^{R}$, $n^{L}$ are calculated in the
$\UV{R}$ and $\UV{L}$ directions respectively. Knowing $n^{R}$ and $n^{L}$ a combined bin index $n$
is calculated. This number is unique for each rectangular $\Delta_{R} \times \Delta_{L}$ bin:
\begin{equation}
\begin{array}{|c|c|c|c|c|}
 \hline
 R  & L & n^{R} & n^{L} & n\\
 \hline
 R_{1} & L_{1} & n^{R}_{1} & n^{L}_{1} & n_{1} \\
 R_{2} & L_{2} & n^{R}_{2} & n^{L}_{2} & n_{2} \\
 \ldots & \ldots & \ldots & \ldots & \ldots\\
 R_{N} & L_{N} & n^{R}_{N} & n^{L}_{N} & n_{N} \\
 \hline
\end{array}
\label{bin_index}
\end{equation}

The next step is sorting the rows of (\ref{bin_index}) so that:
\begin{equation}
n_{\rho(1)} \le n_{\rho(2)} \le \ldots \le n_{\rho(N)}
\end{equation}
where $\rho$ is the permutation implementing the sorting. The resulting table:
\begin{equation*}
\begin{array}{|c|c|c|c|c|}
 \hline
 R  & L & n^{R} & n^{L} & n\\
 \hline
 R_{\rho(1)} & L_{\rho(1)} & n^{R}_{\rho(1)} & n^{L}_{\rho(1)} & n_{\rho(1)} \\
 R_{\rho(2)} & L_{\rho(2)} & n^{R}_{\rho(2)} & n^{L}_{\rho(2)} & n_{\rho(2)} \\
 \ldots & \ldots & \ldots & \ldots & \ldots\\
 R_{\rho(N)} & L_{\rho(N)} & n^{R}_{\rho(N)} & n^{L}_{\rho(N)} & n_{\rho(N)} \\
 \hline
\end{array}
\end{equation*}
can be directly used to calculate the binning. It is sufficient to look at the last column.
The lengths of sequences of identical values of $n$ are the bin counts.
For instance:
\begin{equation*}
\begin{array}{|c|c|c|c|c|}
 \hline
 R  & L & n\\
 \hline
 \ldots & \ldots & \ldots\\
 R_{21} & L_{21} & 3 \\
 R_{3} & L_{3} & 5 \\
 R_{7} & L_{7} & 5 \\
 R_{2} & L_{2} & 7 \\
 R_{11} & L_{11} & 7 \\
 R_{9} & L_{9} & 7 \\
 R_{5} & L_{5} & 9 \\
 R_{12} & L_{12} & 32 \\
 \ldots & \ldots & \ldots\\
 \hline
\end{array}
\end{equation*}
means that there are two hits in bin number $5$,
three hits in bin number $7$ and one hit in bin number $9$. The final result of the 
binning procedure is a list of bins that contain at least one point from $D'$.

\subsection{Measures of performance}
\label{mc}

The simulation, briefly mentioned in the section introduction, produced a total of
almost 9 million registered particle hits - Cartesian $(x , y , z)$ points. 
In order to test the algorithm we used a bootstrap approach and
randomly re-sampled subsets containing $25000$ particle tracks,
this is a typical number of particles tracks (of any transverse momentum) expected simultaneously in one event,
from 
the $2457093$ particle geometries produced by the simulation
(we consider all particle tracks from all events generated by
the simulation as one set). 
Next, from this $25000$ track sample, we removed hits that lie closer than $10 cm$ to the beam line thus eliminating two layers of the pixel barrel and the innermost endcap pixel sensors. The resulting set of registered track positions constitutes $D$ from equation (\ref{d}). 

Finally, from the $25000$ track sample we select a single random charged particle trajectory, the reference
trajectory, with transverse momentum 
greater then $1 GeV$ and at least $7$ registered hits. Using the $x_{c}, y_{c}, \nu$ parameters
of the reference trajectory we transform $D$ into the set $D'$ from \eqref{d_prime}. Next we use the
binning procedure described in section~\ref{binning_procedure}. The result is a list of bins with 
count numbers greater or equal to one. We considered a charged particle track to be present in $D$
if the bin count is greater than or equal to $7$.

Using this re-sampling method it was possible to achieve small uncertainties of the fakes estimation in the plots from section \ref{results} - each plot was calculated using $60000$ data points.
In practice we used a slightly modified but equivalent procedure than
the one described above. It was easier to first select a random track with appropriate properties (transverse momentum, charge, registered number of hits) from the $2457093$ tracks produced by the simulation and then append this reference track to an additional $24999$ randomly chosen trajectories. 
A typical result can be summarized using a table that shows the number of bins with entries greater than 1:
\begin{equation}
\begin{array}{|c|c|}
\hline
\text{bin count} & \text{number of bins} \\
\hline
1 & 1317470 \\
2 & 2595 \\
3 & 15 \\
4 & 1 \\
5 & 1 \\
6 & 0 \\
7 & 0 \\
8 & 0 \\
9 & 0 \\
10 & 1 \\
11 & 0 \\
\hline
\end{array}
\label{example}
\end{equation}
In this example we have chosen a reference track with $10$ registered points in $D$, $p_T=1.488 [GeV / c]$ and $\eta=1.664$. After transforming $D$ into $D'$
all these measurements fall into a single $\Delta R \times \Delta L$ bin that corresponds to the 
helix parameters of the reference trajectory. This is reflected in the table - only $1$ bin has exactly $10$ counts.
Although it is possible to detect helical particle tracks with different $r$ and $z_{0}$ parameters (see Fig. \ref{parametrization}) on the same histogram we consider it very unlikely that two or more helices with the similar
$x_{c}, y_{c}, \nu$ and different $r, z_{0}$ are simultaneously registered in $D$. Ignoring this scenario
we have in (\ref{example}): $1$ true positive result, $1320082$ true negative results, $0$ false positive results, and $0$ false negative results. Note that the procedure from \ref{binning_procedure}
does not return bins with $0$ counts.

More generally, if the bin count in the bin that corresponds to the chosen reference track is greater then or equal to $7$ then:
\begin{itemize}
    \item the number of \emph{true positives} $\text{TP} = 1$, 
    \item the number of \emph{false positives} $\text{FP} = (\text{bins with counts} \ge 7) - 1$, 
    \item the number of \emph{true negatives} $\text{TN} = \text{bins with counts} < 7$,
    \item the number of \emph{false negatives} $\text{FN} = 0$.
\end{itemize}
If the bin count in the bin that corresponds to the chosen track is less $7$ then:
\begin{itemize}
    \item $\text{TP} = 0$, 
    \item $\text{FP} = \text{bins with counts} \ge 7$, 
    \item $\text{TN} = (\text{bins with counts} < 7) - 1$,
    \item $\text{FN} = 1$.
\end{itemize}
Using these quantities it is possible to calculate the efficiency (a distinct quantity from efficiency commonly used in HEP), sensitivity (in HEP referred to as efficiency), specificity and 
predictive value:
\begin{itemize}
    \item $\text{efficiency} = \frac{\text{TP} + \text{TN}}{\text{TP} + \text{FP} + \text{TN} + \text{FN}}$,
    \item $\text{sensitivity} = \frac{\text{TP}}{\text{TP} + \text{FN}} = \frac{\text{TP}}{\text{T}}$,
    \item $\text{specificity} = \frac{\text{TN}}{\text{FP} + \text{TN}}$,
    \item $\text{positive predictive value} = \frac{\text{TP}}{\text{TP} + \text{FP}}$,
    \item $\text{negative predictive value} = \frac{\text{TN}}{\text{FN} + \text{TN}}$.
\end{itemize} 

The procedure described above allows obtaining the statistical characteristics of discussed 
algorithm for a single reference trajectory. This procedure consisting of:
\begin{enumerate}
\item choosing a set containing $25000$ tracks from all available Monte Carlo results, 
\item one reference trajectory from this set is used to calculate $x_{c}$, $y_{c}$, $\nu$ and turn $D$ into $D'$,
\item calculating $\text{TP}$, $\text{TN}$, $\text{TN}$, $\text{FN}$ 
\end{enumerate}
is repeated tens of thousands of times. Using these calculations we can plot the dependence of sensitivity, specificity, predictive value and efficiency as a function of the pseudo-rapidity $\eta$ and transverse momentum $p_{t}$.

\subsection{Results}
\label{results}

For tests three values of $\Delta_{z}$ from (\ref{bin_L}) and $\Delta_{xy}$ from (\ref{bin_R}) were chosen:
\begin{itemize}
    \item $\Delta_{z} = 0.001m$, $\Delta_{xy} = 0.00001m$,
    \item $\Delta_{z} = 0.004m$, $\Delta_{xy} = 0.00004m$,
    \item $\Delta_{z} = 0.016m$, $\Delta_{xy} = 0.00016m$.
\end{itemize}
The bin sizes in $\UV{L}$ and $\UV{R}$ are directly related to $\Delta_{z}$ and $\Delta_{R}$ in equations (\ref{bin_L}) and (\ref{bin_R}).
Plots containing the statistical characteristics of the algorithm are constructed from $60000$ data points in order to arrive at small uncertainties. Separate histograms with bins in either pseudo-rapidity $\eta$ or transverse momentum $p_{t}$ (integrated over the other quantity), containing the numerator
and denominator of the expressions for efficiency, sensitivity, specificity 
and predictive value are created. These histograms are visualised using ROOT~\cite{citeulike:363715}.

\begin{figure}[H]
	\centering
	\includegraphics[width = 0.9 \textwidth]{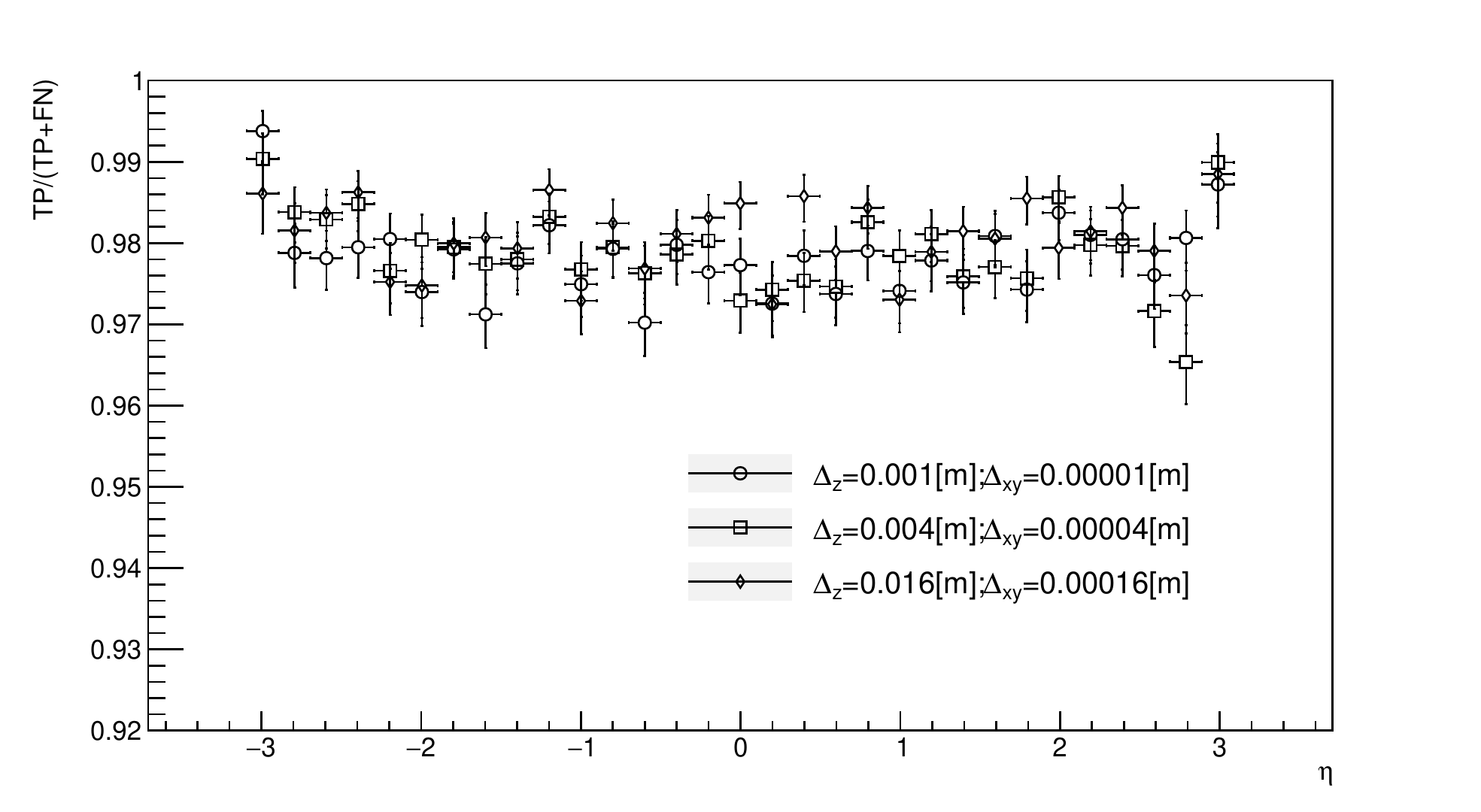} 
	\caption{The sensitivity (efficiency in HEP) $\frac{\text{TP}}{\text{TP} + \text{FN}}$
	as a function of the pseudo-rapidity $\eta$.
	Results for three different pairs of $\Delta_{z}$, $\Delta_{xy}$
	values nearly overlap.
	The uncertainty of the estimate for the sensitivity
	is minuscule.
	}
	\label{sen_eta}
\end{figure}

\begin{figure}[H]
	\centering
	\includegraphics[width = 0.9 \textwidth]{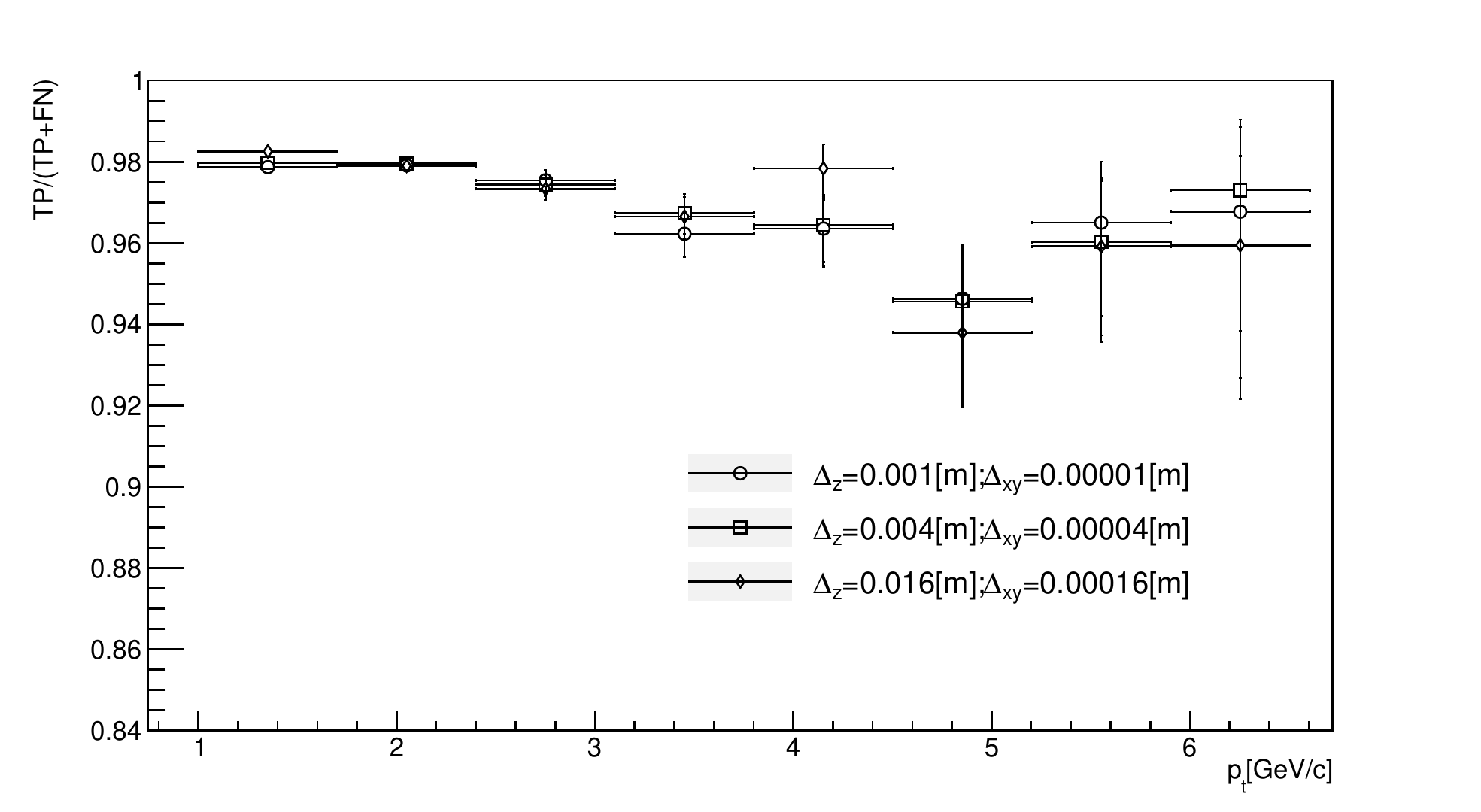} 
	\caption{The sensitivity (efficiency in HEP) $\frac{\text{TP}}{\text{TP} + \text{FN}}$
	as a function of the transverse momentum $p_{t}$.
	Results for three different pairs of $\Delta_{z}$, $\Delta_{xy}$
	values nearly overlap.
	The uncertainty of the estimate for the specificity
	is increasing for higher values of $p_{t}$.
	}
	\label{sen_pt}
\end{figure}

Sensitivity (efficiency in HEP) is a key parameter that specifies the probability that a
helical particle track that is present in the data will be registered  as a positive result (bin count greater or equal to the threshold, helix
detected in bin).  It is shown in Fig.~\ref{sen_eta}
and Fig.~\ref{sen_pt}. For all three pairs of 
$\Delta_{z}$ and $\Delta_{xy}$ values from (\ref{bin_R}) and 
(\ref{bin_L}) a high sensitivity across the entire
considered range of $\eta$ and $p_{t}$ is observed.
For sensitivity as a function of $p_{T}$ uncertainties grow at high values due to quickly falling spectra. Due to the same reason the $\eta$ dependence is dominated by low momentum particles.

\begin{figure}[H]
	\centering
	\includegraphics[width = 0.9 \textwidth]{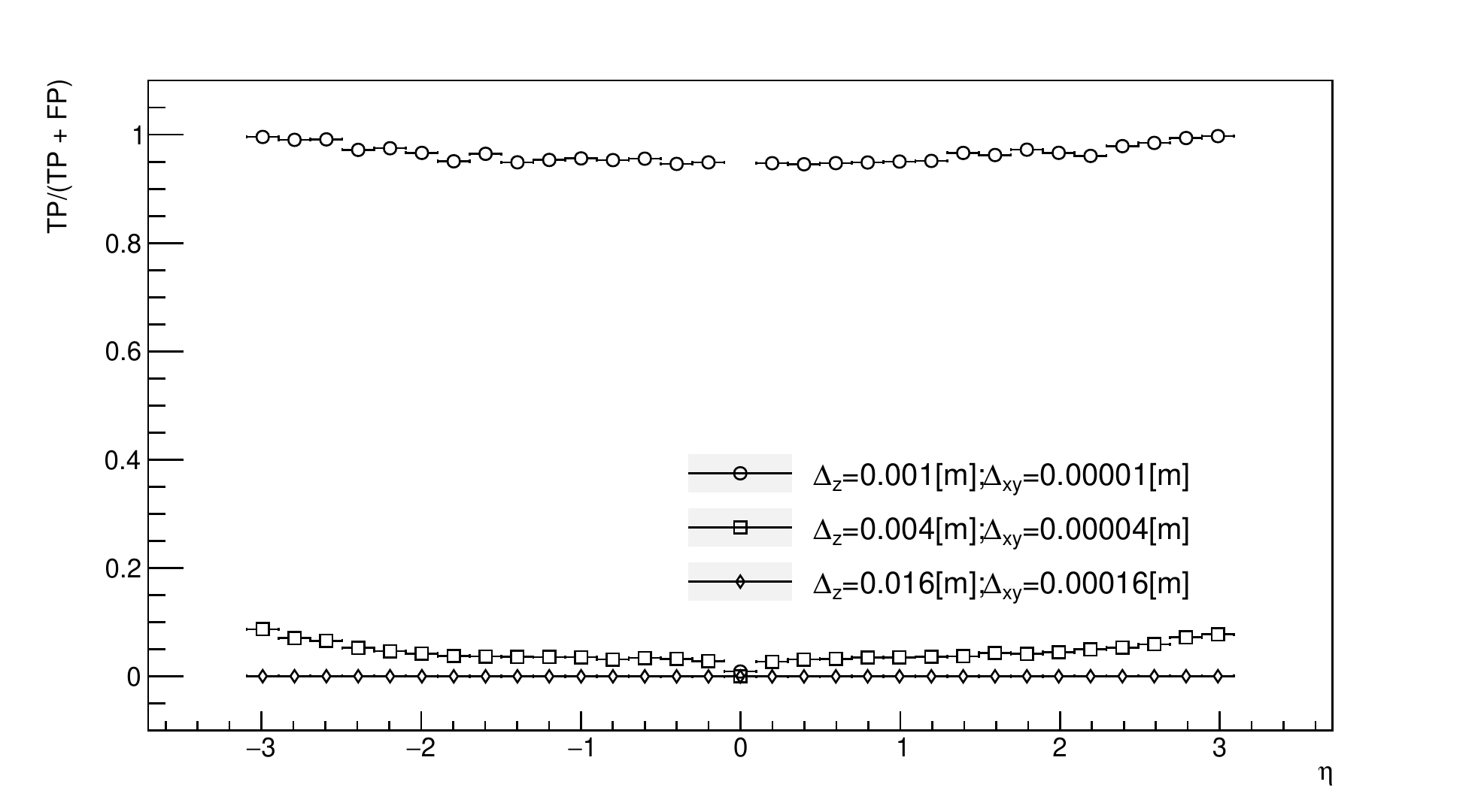} 
	\caption{The positive predictive value $\frac{\text{TP}}{\text{TP} + \text{FP}}$ as a function of the pseudo-rapidity $\eta$.
	A sharp decline in the PPV is observed for the three
	pairs of $\Delta_{z}$, $\Delta_{xy}$ values.
	Additionally a dip in PPV is visible for values of $\eta$ 
	close to $0$.
	The uncertainty of the estimate for the PPV remains small
	in the whole cosidered range of $\eta$.
	}
	\label{ppv_eta}
\end{figure}

\begin{figure}[H]
	\centering
	\includegraphics[width = 0.9 \textwidth]{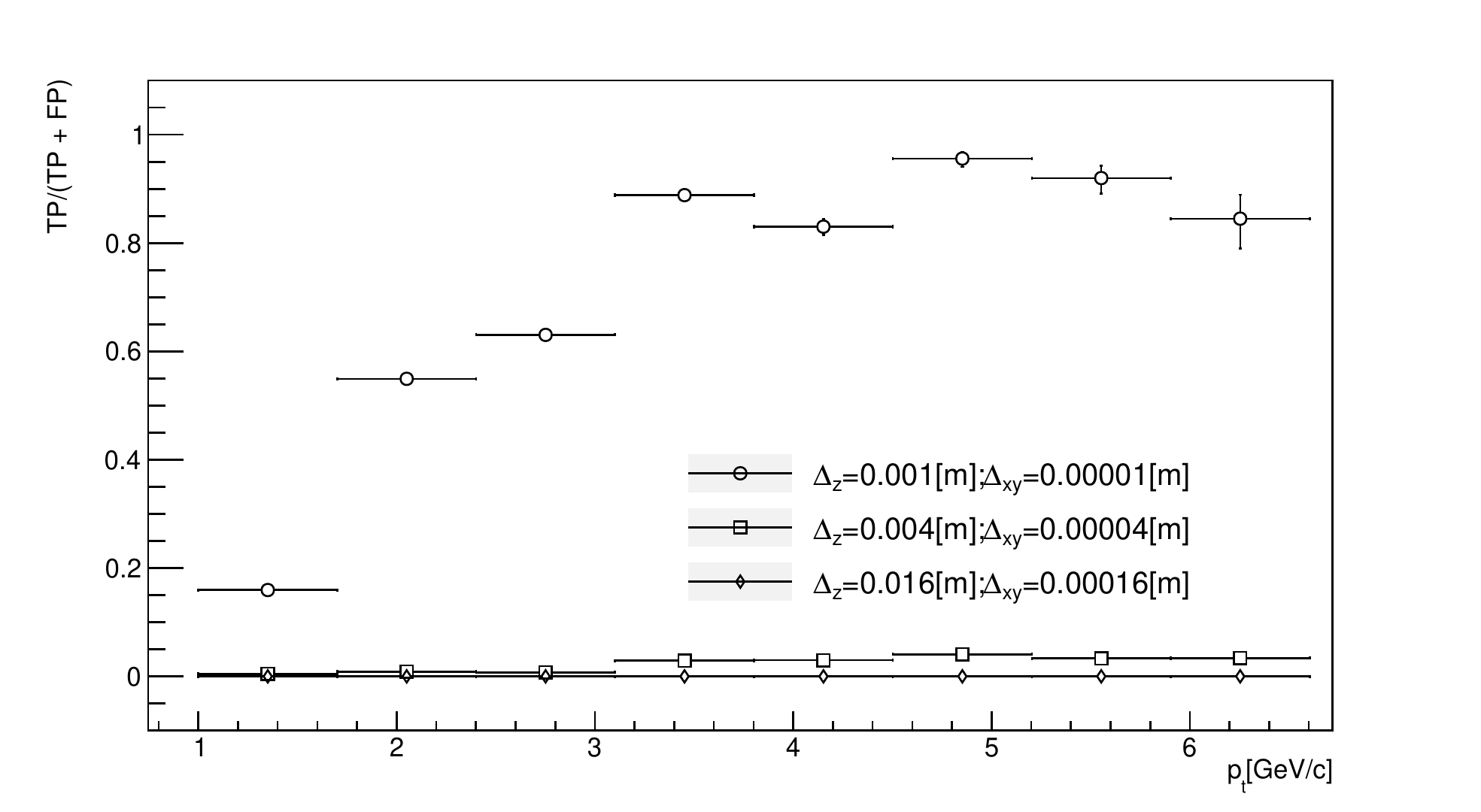} 
	\caption{The positive predictive value $\frac{\text{TP}}{\text{TP} + \text{FP}}$ as a function of the transverse momentum $p_{t}$.
	A sharp decline in the PPV is observed for the three
	pairs of $\Delta_{z}$, $\Delta_{xy}$ values.
	The uncertainty of the estimate for the PPV increases
	for higher values of $p_{t}$.
	}
	\label{ppv_pt}
\end{figure}

Another important parameter is the positive predictive value (PPV).
Values of this parameter 
are related to the probability that given a positive result
a real helix with appropriate parameters is present in the data.
A small PPV means that many helices are artifacts and will need to undergo subsequent processing steps increasing resource consumption, like memory for candidates storage and CPU for fitting. 
The results are plotted on Fig.~\ref{ppv_eta} and Fig.~\ref{ppv_pt}.
On both of these plots a strong dependence of the PPV on $\Delta_{z}$ and $\Delta_{xy}$ values from (\ref{bin_R}) and 
(\ref{bin_L}) can be observed. Furthermore a dip is visible on 
Fig.~\ref{ppv_eta} for values of the pseudo-rapidity close to $0$.
The dip occupies a very small region in $\eta$. Particles 
in this region leave tracks in a thin slice of the detector
that is parallel to the beam line. If the thickness of this 
cross-section is comparable to the $\UV{z}$ resolution of the detector
then the angle of rotation in the unraveling procedure (\ref{u_transformation})
will not be calculated precisely enough. This observation suggests
the necessity to improve the algorithm for this case.
The dependence of the PPV on the transverse momentum
in Fig. \ref{ppv_pt} shows a rising trend and above around $3 GeV/c$
the PPV is close to $1$. That means the proposed algorithm provides a clean sample of of helix candidates at higher $p_t$ while at low $p_T$ the false positives dominate. An impact of appropriate binning is also seen in the plot - too coarse binning results in production of many false positives.

The remaining parameters, the negative predictive value, efficiency, and specificity are less relevant to the operation of the trigger. They are discussed in Appendix \ref{appen}. The values of these
parameters are very close to $1$ for all considered cases and values of $p_{t}$
an $\eta$. 

The results of the performance analysis of the suggested 
algorithm are encouraging. 
Further studies related to optimisation of binning of $D'$ space, algorithm reorganisation for enhanced parallelism, choice of the detector layers are needed to fully prove its applicability.

\subsection{The choice of the step size in the scan for helix parameters}
\label{step_size}

If the proposed algorithm is to be applied to real world data
then it will need to perform a scan in 3D over the helix parameters.
In each iteration the parameters $x_{c}$, $y_{c}$ and $\nu$ need to be changed.
The approach suggested below, based on dimensional analysis, can be used to estimate the optimal choice of this change.
Note that for charged particle tracks originating from the origin the $x_c$ and $y_c$ can be replaced by one variable and thus the scan is over 2D space.

\begin{figure}[H]
	\centering
	\includegraphics[width = 1.0 \textwidth]{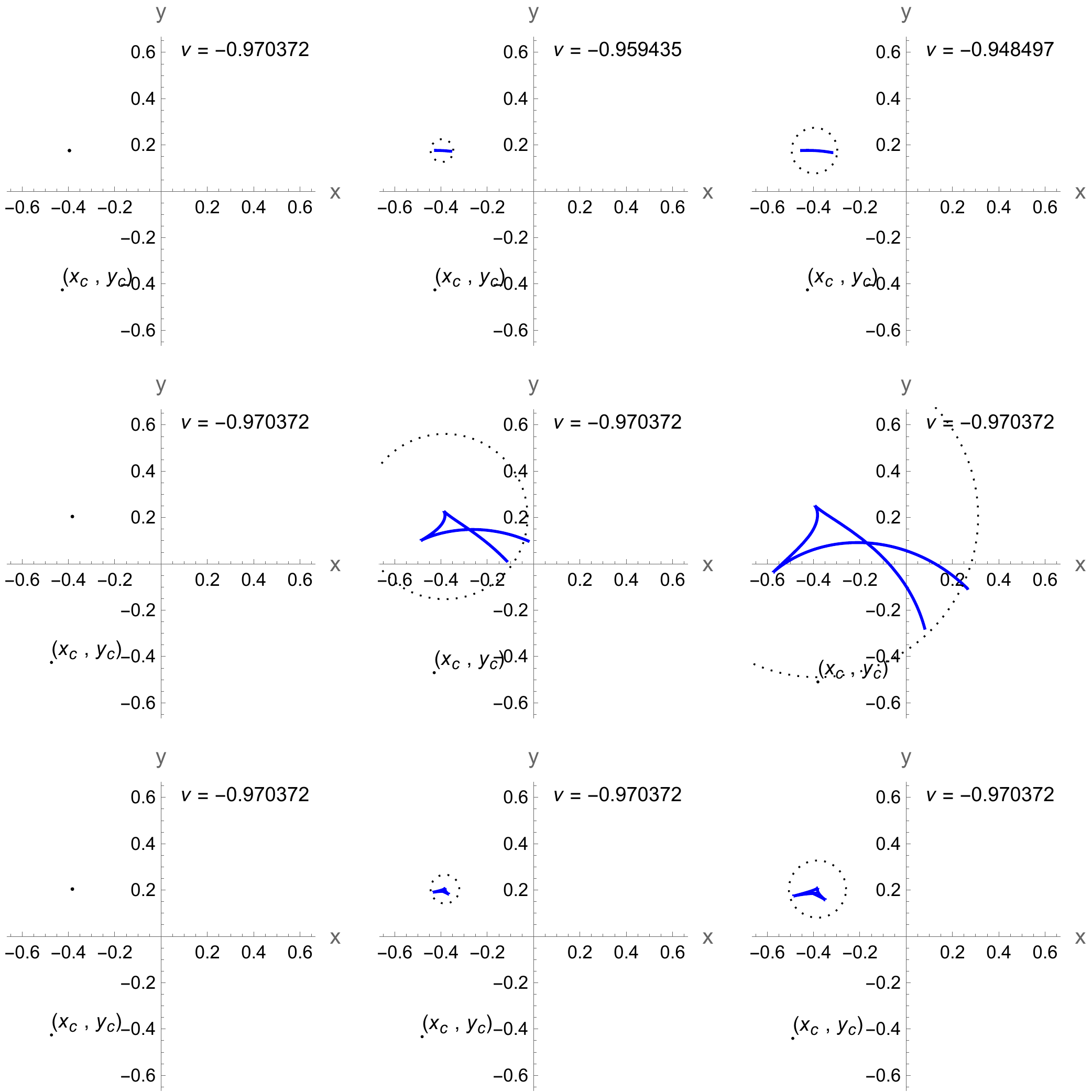} 
	\caption{(color online) Example effect of (\ref{change}) 
	on the size of a helix projection on the $\UV{x} - \UV{y}$ plane. The parameters of the unraveling transformation
	match those of the helix in the leftmost column
	and as a result the projections contain single blue points.
	The top row shows the effect of changing the value of $\nu$.
	Lowering $\nu$ has the effect of gradually 
	"unfurling" the helix. The middle row shows
	the effect of moving the helix center $(x_{c}, y_{c})$
	around the origin. Also in this case a gradual "unfurling"
	of the helix is visible when going from left to right.
	Finally the lowest row shows the effect of moving
	the helix center $x_{c}$, $y_{c}$ away from the origin 
	along the vector $(x_{c} , y_{c})$ and
	the gradual "unfurling" of the helix is shown when going
	from the first to the last column.
	The radius of the dotted circle was calculated using (\ref{runfurl}) with $\alpha$, $\beta$, $\gamma$
	equal to $1.5$, $1.0$, $1.5$ respectively.
	}
	\label{unfurl}
\end{figure}

We consider a situation where the parameters $x_{c}$, $y_{c}$ and $\nu$ match those of a particle track present in the data.
The result is a peak at coordinates $x_{\text{peak}}$, $y_{\text{peak}}$ on the $\UV{x} - \UV{y}$ histogram.
Next we consider a change of parameters:
\begin{align} \nonumber
x_{c}' = x_{c} + \Delta x_{c}, \\ \nonumber
y_{c}' = y_{c} + \Delta y_{c}, \\
\nu' = \nu + \Delta \nu. \label{change}
\end{align}
After (\ref{change}) the straight trajectory in $D'$
will become "unfurled" but will still fit inside a cylinder
with radius $r_{\text{unfurl}}$ around the helix center.
This is illustrated on Fig. \ref{unfurl}.

The cylindrical symmetry of the detector suggests using 
two new parameters
instead
of $\Delta x_{c}$ and $\Delta y_{c}$ in (\ref{change}).
The first new parameter $\Delta \phi_{\text{peak}}$ is 
a change in the azimuthal angle
of the peak relative to the helix center that results from \eqref{change}. The second parameter $\Delta r_{\text{peak}}$ is the change in the distance of the peak from the helix center
that results from \eqref{change}.
Using dimensional arguments we can work out the 
lowest order, linear, expression for a distance
characteristic to the "unfurling" of the helix
as a result of (\ref{change}):
\begin{equation}
    r_{\text{unfurl}} = L_{\text{detector}}
    (\alpha \Delta \nu + \beta \Delta \phi_{\text{peak}} + \gamma \frac{\Delta r_{\text{peak}}}{r_{\text{peak}}} ). 
    \label{runfurl}
\end{equation}
Where $r_{\text{peak}}$ is the distance of the peak to the
helix center, $\alpha$, $\beta$, $\gamma$ are dimensionless constants (in practice close to $1.0$) and we assume that $r_{\text{unfurl}}$ will 
be proportional to the overall detector length $L_{detector}$.
It is important to note that for practical applications the values of $\Delta \phi_{\text{peak}}$,
$\Delta r_{\text{peak}}$ and $r_{\text{peak}}$ in \eqref{runfurl}
can be replaced by quantities that are agnostic to the peak position and related only to the detector geometry. Given the detector radius, only the extreme (maximum or minimum) values of the quantities from \eqref{runfurl} can be considered to arrive at a pessimistic estimate of $r_{\text{unfurl}}$.

Having \eqref{runfurl} the change of parameters \eqref{change}
can be chosen so that $r_{\text{unfurl}}$ does not exceed 
the bin sizes $\Delta R$ and $\Delta L$. This condition 
is sufficient for the main loop of the algorithm not to 
miss any track geometries potentially present in the data. 

\section{Summary and outlook}
\label{sec_summary_and_outlook}

The classification algorithm suggested in this paper
is designed to detect helical particle tracks from charged
particles inside a detector submerged in a uniform magnetic field.
The procedure is performed in a loop. In each iteration
one set of helix parameters is tested
to determine if a corresponding particle track
is present in the data gathered by the detector.
Testing a single set of helix parameters involves
applying a special transformation to all points
registered by the detector and analyzing a
two dimensional histogram for peaks.

The described procedure does not discard
any information and uses all three $(x , y , z)$ components of a particle 
track point registered 
in the detector. Additionally, the main loop of the procedure could be efficiently
parallelized. 
One of the key benefits of the approach 
is that it could be used to look for particle tracks that originate at
large distances from the beam line making it possible to find
particles with longer decay times.

Tests of the statistical performance presented in this paper
are encouraging. The next step of our work will involve 
writing a faster implementation that performs a full scan
of the helix parameters and executes in a parallel manner. It will be interesting 
to study the performance of this new implementation,
compare it to other popular algorithms, and test it
on real data.

\section*{Acknowledgments}

 This work was supported in part by the Polish Ministry of Science and  Higher Education, grant no. DIR/WK/2018/2020/04-1, by the National Science Centre of Poland under grant number UMO-2020/37/B/ST2/01043, and by PL-GRID infrastructure.

\appendix
\section{Appendix}
\label{appen}

\subsection{Negative predictive value}

\begin{figure}[H]
	\centering
	\includegraphics[width = 1.0 \textwidth]{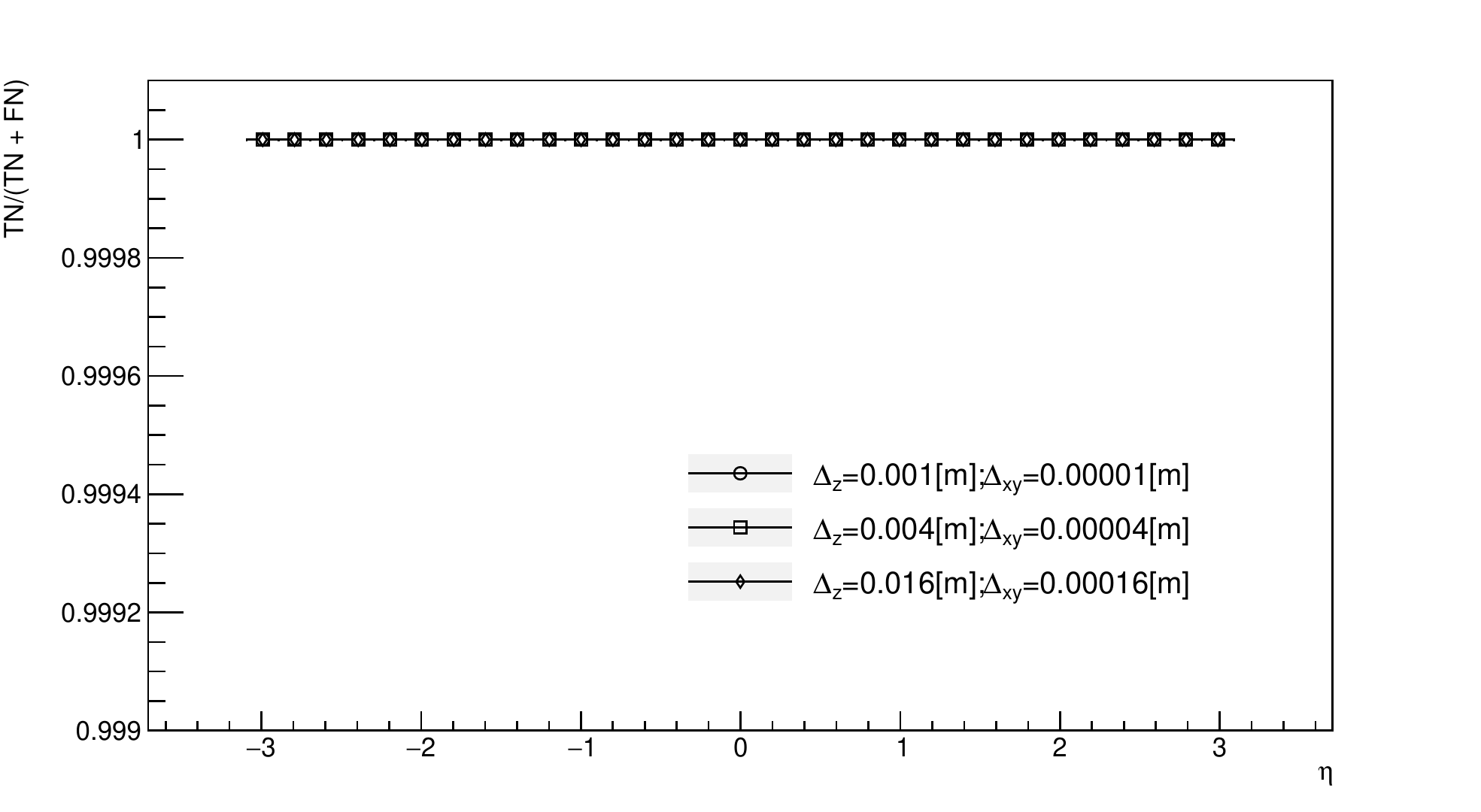} 
	\caption{The negative predictive value $\frac{\text{TN}}{\text{FN} + \text{TN}}$
	as a function of the pseudo-rapidity $\eta$. 
	Results for three different pairs of $\Delta_{z}$, $\Delta_{xy}$
	values overlap. 
	The uncertainty of the estimate for the negative predictive value
	is minuscule.
	}
	\label{npv_eta}
\end{figure}

\begin{figure}[H]
	\centering
	\includegraphics[width = 1.0 \textwidth]{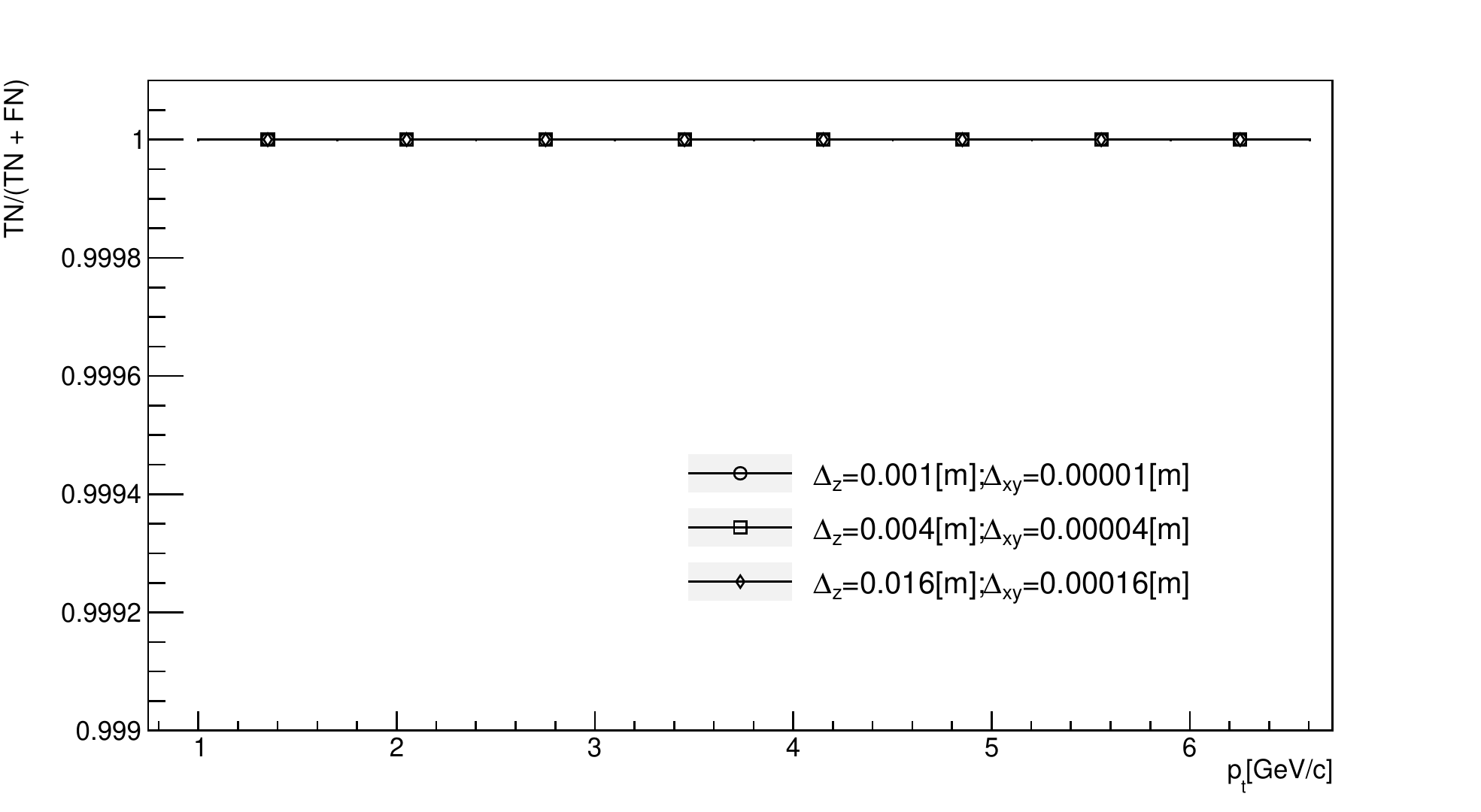} 
	\caption{The negative predictive value $\frac{\text{TN}}{\text{FN} + \text{TN}}$
	as a function of the transverse momentum $p_{t}$.
	Results for three different pairs of $\Delta_{z}$, $\Delta_{xy}$
	values overlap.
	The uncertainty of the estimate for the negative predictive value
	is minuscule.
	}
	\label{npv_pt}
\end{figure}

The negative predictive value is essentially $1$ for all investigated cases,
see Fig.~\ref{npv_eta} and Fig.~\ref{npv_pt}. This is not a surprise
since for each investigated reference track, the vast
majority of bins in $D'$ have the number of entries below the required threshold - see eg. Table (\ref{example}).

\subsection{Efficiency}
\label{efficiency}

\begin{figure}[H]
	\centering
	\includegraphics[width = 1.0 \textwidth]{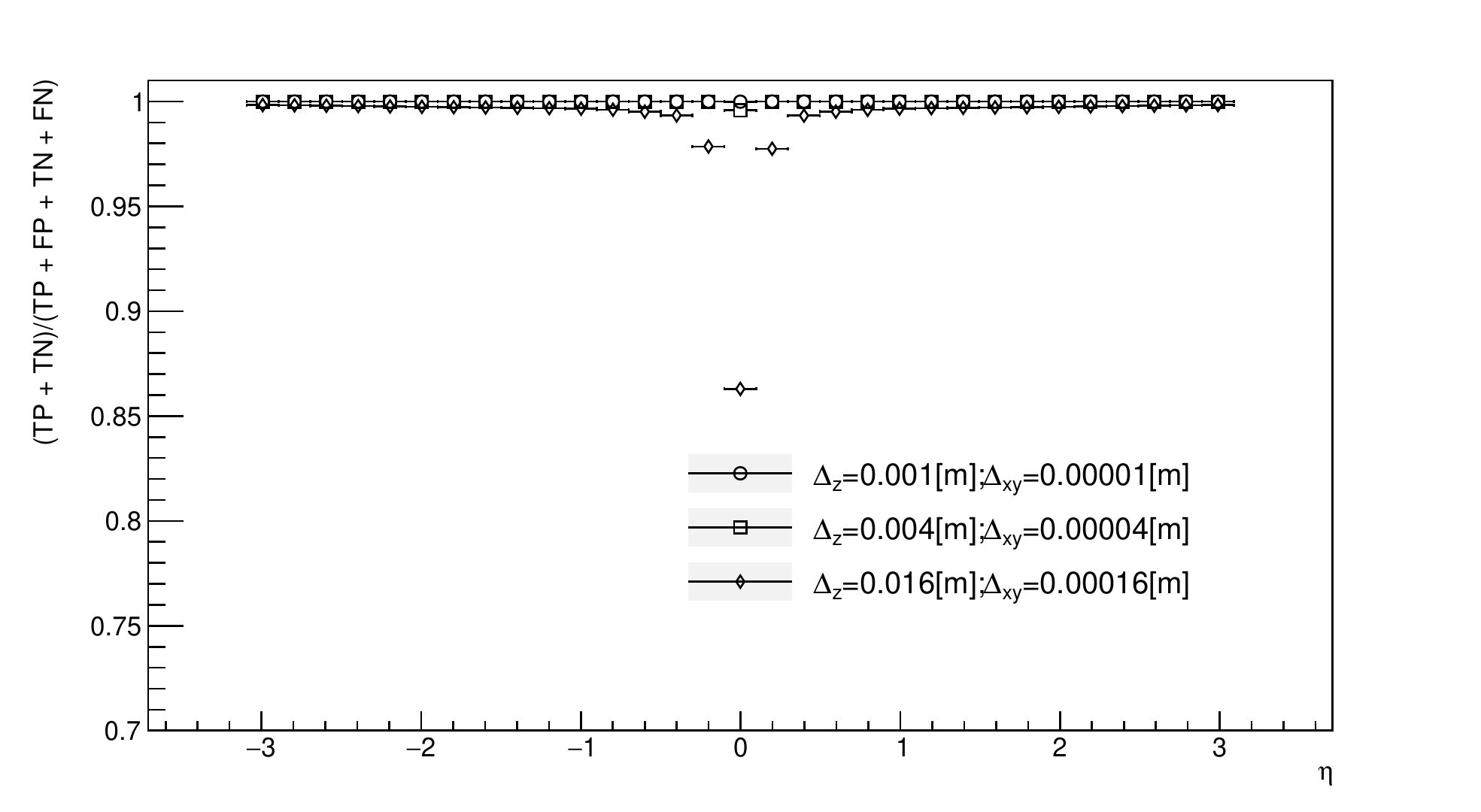} 
	\caption{The efficiency $\frac{\text{TP} + \text{TN}}{\text{TP} + \text{FP} + \text{TN} + \text{FN}}$
	as a function of the pseudo-rapidity $\eta$.
	Results for two different pairs of $\Delta_{z}$, $\Delta_{xy}$
	values overlap. A small dip in efficiency is visible for one pair of
	values.
	The uncertainty of the estimate for the efficiency
	is minuscule.
	}
	\label{eff_eta}
\end{figure}

\begin{figure}[H]
	\centering
	\includegraphics[width = 1.0 \textwidth]{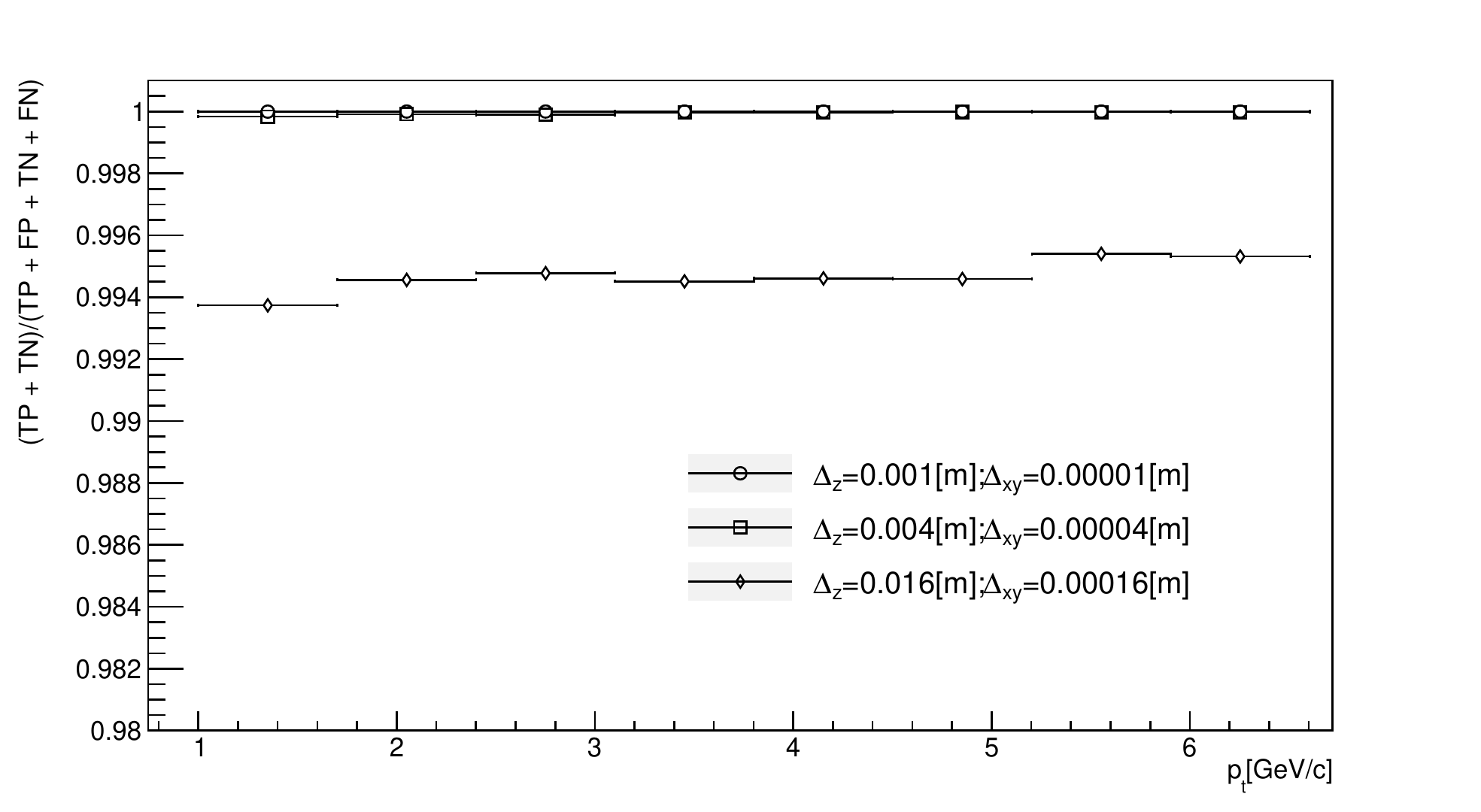} 
	\caption{
	The efficiency $\frac{\text{TP} + \text{TN}}{\text{TP} + \text{FP} + \text{TN} + \text{FN}}$
	as a function of the transverse momentum $p_{t}$.
	Results for three different pairs of $\Delta_{z}$, $\Delta_{xy}$
	values overlap. 
	The uncertainty of the estimate for the efficiency
	is minuscule.
	}
	\label{eff_pt}
\end{figure}

Efficiency (note that this is a different parameter then
the standard HEP definition) is a statistical
measure whose value indicates the probability that the positive results (helical track detected in bin) and negative results (bin count below
threshold, no helix detected in bin) are correct.
The number of true negatives is far greater then the number of true positive
results for a given reference track - see e.g. Table (\ref{example}). For this reason the efficiency is also very close to $1$ as can be seen in Fig. \ref{eff_eta} and Fig. \ref{eff_pt}. Only for the highest considered values of $\Delta_{z}$,
$\Delta_{xy}$ a dip in efficiency is observed in the vicinity of $\eta = 0$. 
The dip occupies a small region in $\eta$. Particles 
in this region leave tracks in a thin cross-section of the detector
that is parallel to the beam line. If the thickness of this 
cross-section is comparable to the $\UV{z}$ resolution of the detector
then the angle of rotation in the unraveling procedure (\ref{u_transformation})
will not be calculated precisely.

\subsection{Specificity}

\begin{figure}[H]
	\centering
	\includegraphics[width = 0.9 \textwidth]{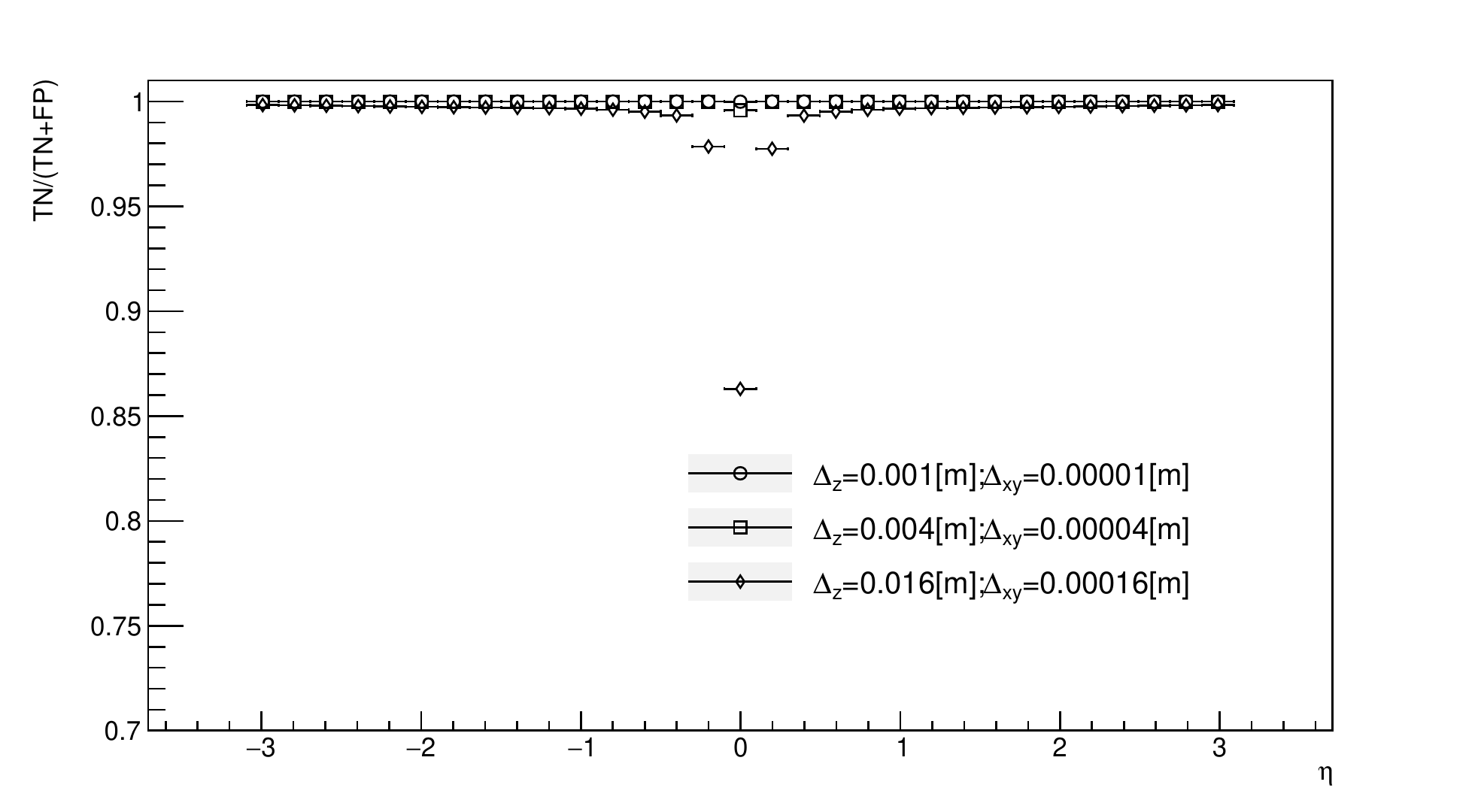} 
	\caption{The specificity $\frac{\text{TN}}{\text{FP} + \text{TN}}$
	as a function of the pseudo-rapidity $\eta$.
	Results for two different pairs of $\Delta_{z}$, $\Delta_{xy}$
	values nearly overlap. A small dip in specificity is visible for one pair of
	values.
	The uncertainty of the estimate for the specificity
	is minuscule.
	}
	\label{spe_eta}
\end{figure}

\begin{figure}[H]
	\centering
	\includegraphics[width = 0.9 \textwidth]{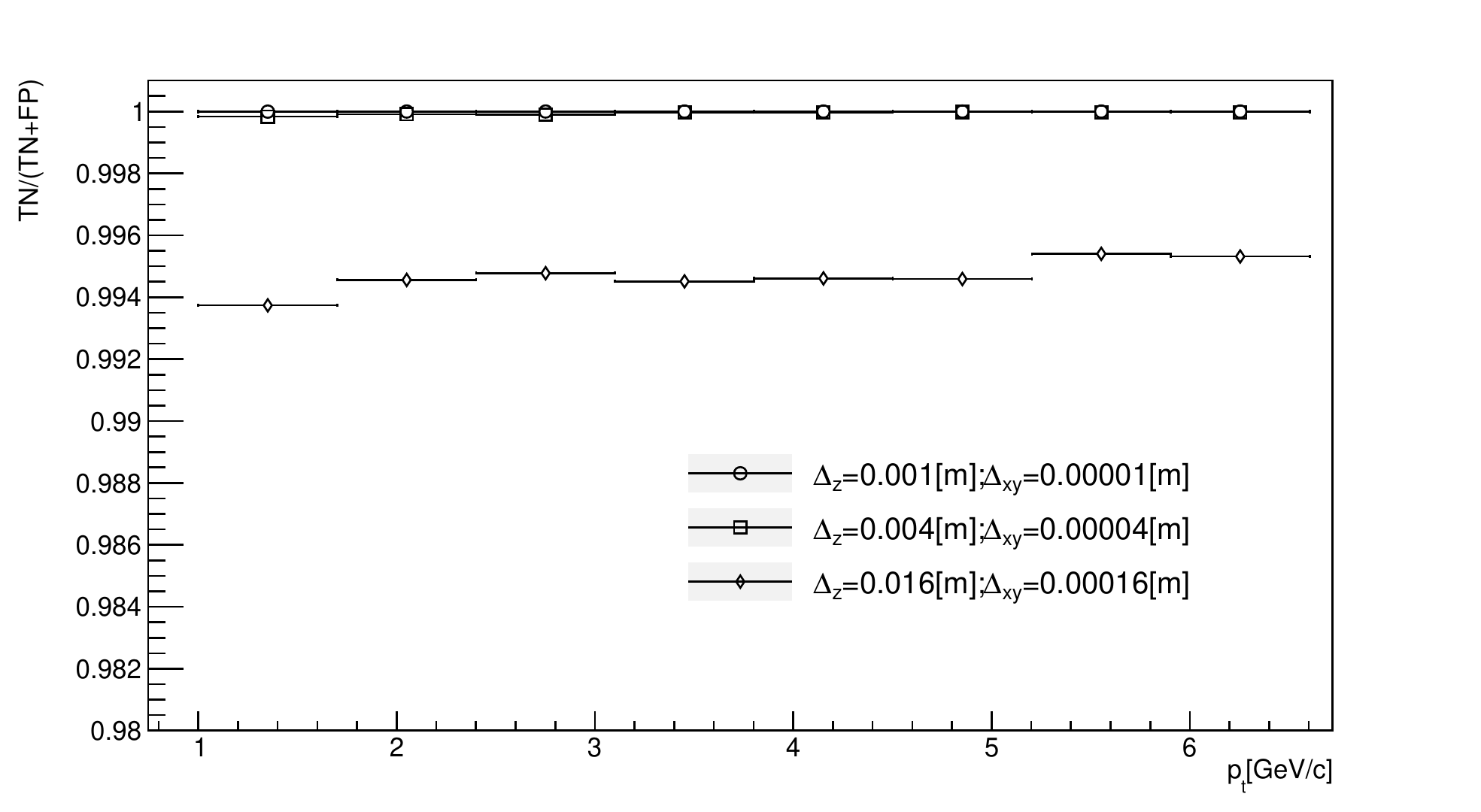} 
	\caption{
	The specificity $\frac{\text{TN}}{\text{FP} + \text{TN}}$
	as a function of the transverse momentum $p_{t}$.
	Results for three different pairs of $\Delta_{z}$, $\Delta_{xy}$
	values nearly overlap.
	The uncertainty of the estimate for the specificity
	is minuscule.
	}
	\label{spe_pt}
\end{figure}

The fraction of true negative results
(bin count below threshold, no helix detected in a bin) among
bins (with non zero count) that should not have registered a positive result is gauged by the specificity. These results are shown on 
Fig. \ref{spe_eta} and Fig. \ref{spe_pt}. All three pairs of 
$\Delta_{z}$ and $\Delta_{xy}$ values from (\ref{bin_R}) and 
(\ref{bin_L}) result in high specificity across the entire
considered range of $\eta$ and $p_{t}$ but a small dip is visible
for values of $\eta$ close to $0$. \
The dip occupies a small region in $\eta$. Particles 
in this region leave tracks in a thin cross-section of the detector
that is parallel to the beam line. If the thickness of this 
cross-section is comparable to the $\UV{z}$ resolution of the detector
then the angle of rotation in the unraveling procedure (\ref{u_transformation})
will not be calculated precisely.

\bibliography{bibl}
\bibliographystyle{ieeetr}

\end{document}